\documentclass[12pt]{article}
\usepackage{amsmath,amssymb}
\usepackage{lmodern}
\usepackage{iftex}
\usepackage[T1]{fontenc}
\usepackage{textcomp} 
\usepackage{graphicx} 
\usepackage{threeparttable}
\usepackage[dvipsnames]{xcolor}
\usepackage[left=2.5cm,right=2.5cm,top=2.5cm,bottom=2.5cm]{geometry}
\usepackage{longtable,booktabs,array}
\usepackage{calc} 
\usepackage{caption}
\usepackage{csquotes}
\usepackage{hyperref}
\hypersetup{
    colorlinks=true,
    linkcolor=blue,
    citecolor=blue,
    filecolor=magenta,      
    urlcolor=blue,
    pdftitle={The Fitness of the European Labour Force},
    pdfpagemode=FullScreen,
    }
\usepackage{authblk}
\usepackage{comment}

\usepackage{setspace}

\usepackage{natbib} 
\setcitestyle{authoryear,open={(},close={)}} 

\begin{document}

\begin{titlepage}

\title{Structural Change, Employment, and Inequality in Europe: an Economic Complexity Approach}
\author[1,2,3]{Bernardo Caldarola\thanks{Corresponding author. Email: \url{bernardo.caldarola@ec.europa.eu}}}
\author[1]{Dario Mazzilli}
\author[1]{Aurelio Patelli}
\author[1]{Angelica Sbardella}

\affil[1]{European Commission, Joint Research Center, Seville, Spain}
\affil[2]{Enrico Fermi Research Center, Rome, Italy}
\affil[3]{UNU-MERIT and Maastricht University, Maastricht, the Netherlands}

\maketitle
\footnotesize
\begin{abstract}
\singlespacing

Structural change consists of industrial diversification towards more productive, capability-intensive activities, bearing inherent links with job creation and income distribution. This paper investigates the consequences of structural change –- defined in terms of labour shifts towards more complex industries –- on employment growth, wage inequality, and the functional distribution of income between labour and capital. The analysis is conducted for European countries using data on granular sector-level employment shares over the period 2010-2018. Structural change is described using a country-level measure that captures the movement of workers towards more complex industries. The findings indicate that the complexity-based measure of structural change is associated negatively with employment growth. However, it is also associated with lower income inequality: as countries move to more complex industries, they drop the least complex ones, so the (low-paid) jobs in the least complex sectors disappear. Finally, structural change predicts a higher labour ratio of the economy; however, this is likely to be due to movement of the workforce into higher-wage, more complex sectors rather than to job creation.

\end{abstract}
\begin{flushleft}
\textbf{Keywords:}  Structural Change, Economic Complexity, Wage Inequality, Labour Share, Employment Growth \linebreak
\textbf{JEL Codes:} D63, E24, J31, O11, O52 
\end{flushleft}
\end{titlepage}

\section{Introduction}\label{sec:intro}

Structural change describes the process of reallocation of economic activity across economic sectors \citep{Lewis1954, Syrquin1988, matsuyama2008} with heterogeneous levels of productivity, technological content, and labour requirements \citep{baumol1967macroeconomics} which drives long-term economic growth \citep{Herrendorf2014, Herrendorf2022}. Commonly, structural change is quantified by changes in the relative importance of industries -- measured by their shares in total employment or output. This process has been shown to rest on the underlying accumulation of increasingly sophisticated productive capabilities \citep{verspagen1991new, abramovitz1995elements,cimoli_industrial_2009}, which allows economies to innovate and diversify through the recombination of existing knowledge \citep{Hausmann2009, Tacchella2012, rodrik2013structural, vandam2022, inoua2023, teixeira_scientific_2026}.

However, structural change is neither a uniform nor an automatic process. Different varieties of structural transformation may lead to distinct outcomes depending on the sectors into which economies diversify and the institutional contexts in which these transformations unfold, according to which its gains are distributed across workers and between labour and capital. This is particularly relevant in advanced economies, where structural change has increasingly taken the form of deindustrialisation, tertiarisation, and the expansion of knowledge-intensive sectors rather than the classical shift from agriculture to manufacturing \citep{galbraith2011inequality, Herrendorf2014, rodrik2016premature}. Against this backdrop, this paper focuses on one specific aspect of this broader process: how structural change, conceptualised as the reallocation of labour towards more complex and capability-intensive sectors, is associated with employment creation and income distribution in European economies.

In his seminal work, \citet{Kuznets1955} argued that inequality would initially rise and then decline as economies transition from traditional to manufacturing sectors. In this framework, income disparities arise from productivity differentials across sectors: as long as low and high productivity sectors coexist, aggregate inequality reflects the income gap between them. Inequality would initially rise as a modern sectors start to grow, but as sectoral reallocation converges towards the higher-productivity sectors, inequality is predicted to decline. Although this inverted-U shape relationship between development and inequality proposed by Kuznets has been widely debated \citep{atkinson2006panel, hellier2013inequality, galbraith2025kuznets}, the broader insight that sectoral reallocation and income distribution are structurally linked remains influential. The multiplicity of definitions of both “structural change” and “inequality”, together with the diversity of empirical patterns observed across countries and historical periods, has prevented the emergence of a unified consensus regarding the direction and magnitude of this relationship \citep{cosentino_structural_2023,Forster2015,kanbur_income_2000}. Moreover, several mechanisms are at play along with sectoral changes, with an effect of income distribution and employment creation. These include globalisation, technological change, labour market institutions, political processes, tax regimes, and demographics \citep{Forster2015}. 
Nevertheless, the extant literature on this topic has not built a direct link between the accumulation of (increasingly rare) capabilities as sectoral changes unfold and the distribution of income. 

In this paper, we contribute to this debate by investigating the labour consequences of knowledge recombination towards more sophisticated capabilities. To this aim, we use a definition of structural change that moves beyond measures based on productivity or sectoral shares, and builds explicitly on capabilities accumulation a fundamental notion of structural change. Our definition and empirical operationalisation of structural change pins down the movement of labour to industries that require more sophisticated capabilities using the Economic Complexity framework \citep{Hausmann2009, Tacchella2012}. First, we construct an employment-based measure of Economic Fitness which quantifies the diversification of industries specialisation of a country, weighted by the sophistication of such industries. Second, we separate changes in labour-weighted Fitness between i) the reallocation of labour towards more sophisticated and capability-intensive sectors, and ii) changes in industry sophistication due to changes in the capabilities required by the industry itself. We relate the former -- the \textit{structural change} component -- to employment growth, wage inequalities, and the functional distribution of income between labour and capital. The proposed approach offers a perspective on structural change and inequalities that links directly the process of capability accumulation that underlies structural transformations to the patterns in employment creation and labour income distribution associated to it. 

Understanding these dynamics is particularly relevant in light of the profound transformations observed in European and other high-income economies over recent decades, which have been accompanied by rising within-country inequality, declining employment growth, and a falling labour share of income.  Within-country inequality has been on the rise in Europe and in other high-income countries \citep{Picketty2014, Picketty2018}, with increasing between-industry wage dispersion \citep{Haltiwanger2024}. Increasing wage inequalities have been accompanied by a decline in the labour share of income across European countries and the US alike \citep{jorgenson2011structural, karabarbounis2014global, stockhammer2017determinants, Karabarbounis2024}, redistributing wealth away from workers towards capital owners. At the same time, most high-income countries have de-industrialised during the last wave of ICT-led globalisation, leading to the substitution of manufacturing with jobs in complex, knowledge-intensive industries, and contributing to increasing inequalities \citep{cirillo2018job, Antonelli2020}. 

The mechanisms underlying these trends has been widely explored in the empirical literature. The Skill-Biased Technological Change (SBTC) literature \citep{Autor2003, Autor2022, Acemoglu2022} has examined the impact of technological change, the main driver of structural change, on employment and on the distribution of earnings, suggesting that recent waves of technological change, especially ICT capital goods, tend to complement high-skilled workers in knowledge-intensive industries, leaving low-skilled workers behind. Later this literature has argued that digital technology-driven transformations may affect not only the relative demand for skilled labour, but also the occupational structure itself, giving rise to increasing job polarisation, that is, to the disappearance of middle-wage jobs \citep{autor_polarization_2006, Goos2009} which are more easily replaced by capital \citep{acemoglu_automation_2026}.
Additionally, offshoring, the expansion of global value chains, and import competition from lower-wage countries have been seen as contributing to reducing the demand for low- and middle-skilled labour in advanced economies, thereby reinforcing wage inequality and labour market polarisation \citep{feenstra1996globalization, autor2013china, timmer2014slicing}. Alongside these market-based explanations, the institutional literature interprets the distributive outcomes associated with structural change not as the mechanical result of technology and trade alone, but also of political and institutional changes that shape wage-setting mechanisms and the division of income between labour and capital. In this perspective, rising inequality and the decline in the labour share reflect not only shifts in labour demand, but also the erosion of workers’ bargaining power through declining unionisation, labour market deregulation, weaker labour standards, and welfare state retrenchment \citep{mishel2021identifying,stansbury2020declining,stockhammer2012have}.

Another crucial feature of structural change is its linkage with employment creation. As economies transform in their productive structures, new jobs are created while some others are made redundant \citep{Schumpeter1934}. As the employment effects of structural change depend on the labour intensity of expanding sectors, structural upgrading may be associated with lower employment growth even in the presence of rising productivity \citep{kucera2012structure}. On the one hand, the current wave of automation may not reduce total employment \citep{Dauth2021}, but may induce labour market adjustments at the disadvantage of low-skilled workers, whose share over total workers is reduced \citep{Graetz2018}. On the other hand, ICTs may have been among the culprits of jobless growth \citep{Brynjolfsson2011, Frey2017} and contributed to the shrinking of the number of middle-skill jobs \citep{Goos2009, Goos2014} due to the lower labour requirements of ICT intensive industries.        

In order to identify variety and uniqueness of capabilities required to achieve industrial upgrading we employ an Economic Complexity (EC) perspective. Economic Complexity metrics aim at quantifying the hierarchical capability structure of countries by looking at their productive basket, preserving information on their specialisation and diversification patterns rather than by simply aggregating their economic output \citep{Hausmann2009, Tacchella2012, diodato_handbook_2024}. By the same token, EC measures allow to identify the capability requirements \citep{Loturco2022} of individual products, industries, or technologies. Given the strong link of EC metrics with economic diversification, and in particular of the Economic Fitness and Complexity (EFC) approach \citep{Tacchella2012}, they have proved to be a useful tool to describe a long-term dynamic such a structural change \citep{Freire2021, Castaneda2022, Mcnerney2025}, being also highly predictive of GDP growth \citep{Tacchella2018}. 

Following the analogy between structural change and complexity, the latter can be used to explain the inequality trends that result from changes in the sectoral composition of the economy. The literature on complexity and inequalities \citep{Sbardella2017, Hartmann2017, Hartmann2022} indicates that more complex activities are better remunerated, and that higher complexity of the productive structure is associated to lower inequalities at the country level \citep{Sbardella2017, Hartmann2017, hartmann2024economic, aufiero2024mapping}. However, this relationship may present non-linearites and also reflect cross-country differences in institutional settings, including the inclusiveness of labour market and welfare institutions \citep{Hartmann2017, Sbardella2017, napolitano2022green, hartmann2024economic}. Looking at the relationship between complexity and employment, \citet{Adam2023} finds that higher complexity is linked to job creation. Where existing jobs are made redundant, it has been shown that re-employment is easier in presence of related industrial variety at the regional level \citep{HaneWeijman2018}.  

We measure the relative importance of finely disaggregated industries (NACE 4  digits) using data on industrial employment in European countries between 2010 and 2018\footnote{Similar efforts to compute Economic Complexity indices for countries or regions have been performed, among others, by \citet{Sbardella2017, caldarola2022structural, Pinheiro2022}} obtained from Eurostat’s Structural Business Survey data and compute a measure of Economic Fitness weighted by labour shares. Using data on industrial employment allows us to include also non-tradeable activities, most of which are in the service sector.\footnote{\citet{mishra2020} have also been able to include services activities in the computation of Economic Fitness, however limited to tradable ones.} This is particularly relevant considering that European economies have heavily shifted their productive structure towards the tertiary industries along with the process of globalisation. Secondly, the focus on employment represents a crucial element in the conceptualisation of structural change as the change in relative shares of the economy. 
Data on the individual and functional distribution of income are obtained, respectively, from ILO and ARDECO; the latter is also used to measure employment growth in the observed period. First, we identify patterns of industrial specialisation by validating the country-industry industrial employment matrix using a null model approach -- the Bipartite Weighted Configuration Model  \citep{bruno_inferring_2023} -- that overcomes some limitations imposed by Balassa's (\citeyear{Balassa1965}) Revealed Comparative Advantage index. Secondly, we introduce a measure of labour-weighted Fitness, which sums up the complexity of industries weighted by their employment share. This measure can be decomposed in such a way as to identify the contribution to changes in labour-weighted Fitness coming from the movement of labour towards more complex industries. We identify such component as the one linked to structural change from a labour viewpoint –- the structural change component. Thirdly, we link the structural change component to a number of economic outcomes at the country level: i) employment growth, ii) wage inequality, and iii) the functional distribution of income (labour share of the economy). In this way, we analyse the distributional consequences of structural transformation -- framed as the movement of labour to more complex industries -- and its job-creating potential. 

Our analysis of the specialisation patterns in European countries reveals that the most diversified countries have progressively moved away from specialisation in low-complexity, labour-intensive industries. The results of a fixed effect panel regressions indicate that our structural change measure is negatively associated with employment growth, corroborating the evidence that highly complex industries have lower labour requirements. However, it is also associated with lower income inequality –- measured in terms of the ratio of average wages in the ninth and first deciles of the wage distribution. As countries move to more complex industries, they drop the least complex ones, so the (low-paid) jobs in the least complex sectors disappear, making the first decile of the salary distribution go up. Finally, structural change is associated with a higher labour share of the economy, likely mirroring a reallocation of labour towards complex and higher-paying  industries rather than a employment expansion. Interestingly, the structural change component is the only one to be significant in the regressions, while the increase in within-sector complexity, and the labour-weighted Fitness of countries do not explain our outcome variables. Results are robust to a battery of robustness checks, including leave-one-out regressions and a specification where the structural change component of the labour-weighted Fitness is replaced by a similar component obtained decomposing a labour-weighted entropy measure, that takes into account the breath of sectoral specialisation but not its complexity. These findings imply the need for industrial, innovation, and labour-market policies capable of supporting transitions towards more complex activities while preserving job creation capacity and ensuring that the gains from structural transformation are distributed in an inclusive way. 

The remainder of this paper is organised as follows. In Section \ref{sec:framework} we delve into the Economic Fitness and Complexity framework, and we introduce a new measure, the Labour-Weighted Fitness (LWF) the variation of which can be decomposed to extract a structural change component. In Section \ref{sec:data} we outline the data sources used in the analysis, which in turn is described in the following Section (\ref{sec:methods}). Section \ref{sec:results} illustrates the main results of the analysis, and Section \ref{sec:conclusions} concludes.

\section{Structural change and Economic Complexity}\label{sec:framework}

In this section, we outline the analytical framework used to derive a measure of structural change based on the reallocation of labour to complex industries. First, we will define country Fitness and industry complexity according to the EFC framework. Second, we introduce a new measure -- the Labour Weighted Fitness -- from which we derive the structural change component of interest. 

\subsection{The Economic Complexity framework}\label{sec:struchange}

The Economic Complexity framework builds on earlier evolutionary and institutional traditions that emphasise the role of productive capabilities, sectoral composition, and cumulative learning in shaping development trajectories \citep{Cimoli1995, Hirschman1958, Teece1994}. At its core lies the idea that the production of goods and services requires combinations of highly specific and complementary underlying capabilities \citep{abramovitz1995elements,hausmann2003economic}, such as skills, technologies, organisational routines and institutions. Because these capabilities cannot be measured directly, the Economic Complexity approach infers their distribution from observable patterns of specialisation across economic activities. By examining the industries or products in which countries display a comparative advantage, it becomes possible to recover information about the underlying structure of capabilities embedded in their economies \citep{Hausmann2006,Hausmann2009,Tacchella2012}. This perspective shifts the focus from aggregate outcomes, such as GDP, to the composition of productive activity, asking in which industries do countries specialise and what does this reveal about their underlying capabilities. To do so, Economic Complexity methods represent specialisation profiles as networks linking countries to the activities in which they display a comparative advantage, and then exploits the structural information embedded in these networks to construct ad hoc metrics grounded in complex systems theory. By leveraging highly disaggregated data and tools from network science and complex systems analysis, the Economic Complexity framework provides a coherent empirical approach to study diversification, technological upgrading, and structural transformation, complementing more traditional (and aggregate) analyses. In particular, EC methods have contributed \textquote{to the study of structural change by addressing questions related to how the diversification of countries and the patterns of trade specialization affect and are affected by economic growth, and how diversification and trade specialization change over time, and the determinants of such change} \citep[p~214]{Freire2021}. 

In order to measure the Fitness of countries and the Complexity of European industries, we rely on the Fitness and Complexity algorithm proposed by \citet{Tacchella2012}:

\begin{equation}
    \left\{\begin{array}{r@{}l@{\qquad}l}
	\widetilde{F}_c^{(n)}=\sum_i M_{ci} Q_i^{(n-1)} &\qquad & F_c^{(n)}=\frac{\widetilde{F}_c^{(n)}}{\langle\widetilde{F}_c^{(n)}\rangle_g} \\ \\
	\widetilde{Q}_i^{(n)}=\frac{1}{\sum_{c} M_{ci} \frac{1}{F_{c}^{(n)}}}  &\qquad & Q_i^{(n)}=\frac{\widetilde{Q}_i^{(n)}}{\langle\widetilde{Q}_i^{(n)}\rangle_a}
    \end{array}\right.
    \label{eq:fitness}
\end{equation}

with $Q^{(0)}_i = 1$. The algorithm takes as input a binary country-industry employment specialisation matrix $M_{c,i}$ constructed by cross-tabulating industrial employment data across European countries. The matrix is binarised according to the industrial specialisation profile of each country and indicating whether country $c$ displays a comparative advantage in industry $i$.\footnote{In this work, the binary specialisation matrix $M_{c,i}$ -- which identifies industries in which countries are specialised -- is obtained using an entropy maximising null model (the Bipartite Weighted Configuration Model by \citet{bruno_inferring_2023}) and implementing a method to ensure the comparability of Fitness and Complexity values over time. A more in-depth explanation of the construction of the matrix are explained in Section \ref{sec:biwcm}. In Appendix \ref{sec:ICA} we detail the advantages of using an entropy-maximising null model to measure comparative advantages, over traditional methods such as the Revealed Comparative Advantage \citep{Balassa1965}.} 

Starting from this matrix, the iterative formula in Equation \ref{eq:fitness} jointly determines the values of country Fitness (first row of Equation \ref{eq:fitness}) and industry complexity (second row of Equation  \ref{eq:fitness}). The intuition of the algorithm is straightforward: a country is assigned a high Fitness if it is specialised in many complex industries. Conversely, an industry is assigned a high Complexity if it hard to find, and when it is found, it appears in countries that are highly diversified. More formally, the Fitness $F_c$ of country $c$ at each time period corresponds to the weighted sum of the Complexities $Q_i$ of the industries $i=1,..., N$ in which the country holds a comparative advantage, as identified by the binary specialisation matrix $M_{c,i}$. In turn, the complexity $Q_i$ of industry $i$ corresponds to the inverse of its ubiquity weighted by the inverse of the Fitness of the countries which have that industry in their portfolio. More complex industries, therefore, are the least ubiquitous ones -- proxying for the rareness of the capabilities required by those -- considering also how 'fit' are the countries that hold a specialisation in that industry. The algorithm is iterated $n$ times, until it converges. At each step, Fitness and Complexity values are normalised by dividing them by the respective average values. A feature that makes Fitness a suitable indicator to measure structural change as the process of capabilities accumulation structure is that it considers the cumulative nature of capabilities by considering explicitly the diversification of countries (weighted by the complexity of their industries). This feature is absent in other methods, such as the Economic Complexity Index (ECI) \citep{Hausmann2009}, which do not consider diversification when measuring the complexity of countries' productive structures \citep{Cristelli2013, Mcnerney2025, neffke2026economic}. This property comes in handy to decompose the Economic Fitness measure and isolate its structural change component, as explained in the next subsection. 
 
\subsection{Labour-Weighted Fitness and decomposition}\label{sec:lwf}

The starting point to construct a measure of structural change that captures the reallocation of labour across industries is to manipulate the Fitness equation (the first row of Equation \ref{eq:fitness}) in such a way as to weight the sum of the complexities $Q_i$ by their labour share. Thus, for $N$ industries, the Labour-Weighted Fitness can be expressed as: 

\begin{equation}
 LWF_t = \sum_{i=1}^N \theta_{i,t}Q_{i,t} 
 \label{eq:lwf}
\end{equation}

Where $\theta_{i,t}$ is the employment share of industry $i$ at time $t$ for any given country. The resulting LWF is therefore an index of the Fitness of countries, where more weight is given to industries that are not only complex, but which also absorb a larger employment share, indicating the presence of knowledge-intensive capabilities across a larger portion of the workforce. To extract our measure of structural change, we now propose a structural decomposition of LWF inspired by the productivity decompositions adopted in the structuralist literature \citep{McMillan2014, McMillan2017a, Vries2015}, which originates from the original work of \citet{Fabricant1942}. We decompose $\Delta LWF_t$ -- the variation of LWF between time $t$ and $t-k$, with $0\leq k \leq T$ -- in within- and between-sector components as follows: 

\begin{align}
& \Delta LWF_t = LWF_t - LWF_{t-k} = \label{eq:dec1} \\ 
& (=) \ \sum_{i=1}^N \theta_{i,t}Q_{i,t} - \sum_{i=1}^N \theta_{i,t-k}Q_{i,t - k} = \label{eq:dec2} \\ 
& (=) \ \sum_{i=1}^N \theta_{i,t}Q_{i,t} - \sum_{i=1}^N \theta_{i,t-k}Q_{i,t - k} \color{OliveGreen}{\, + \sum_{i=1}^N \theta_{i,t-k}Q_{i,t} - \sum_{i=1}^N \theta_{i,t-k}Q_{i,t}} = \label{eq:dec3}\\
& (=) \ \underbrace{\sum_{i=1}^N \theta_{i, t-k}\Delta Q_{it}}_\text{WITHIN TERM} + \underbrace{\sum_{i=1}^N Q_{it}\Delta \theta_{i, t}}_\text{BETWEEN TERM}. \label{eq:dec4}
\end{align}

Therefore, rearranging terms, Equation  \ref{eq:dec1} yields the decomposition in the two distinct terms described in Equation  \ref{eq:dec4}: 

\begin{itemize}
    \item $\sum_{i=1}^N \theta_{i, t-k}\Delta Q_{it}$ is the \textbf{within-sector term of structural change} (the contribution of increasing complexity at the sectoral level to the change in labour-weighted Fitness);
    \item $\sum_{i=1}^N Q_{it}\Delta \theta_{i, t}$ is the \textbf{between-sector term of structural change} (the contribution of labour shifts from low to high complexity sectors to the change in labour-weighted Fitness)
\end{itemize}

The between-component, given by $\sum_{i=1}^N Q_{it}\Delta \theta_{i, t}$, identifies the measure of structural change of interest. This will be used to explain the observed trends in wage inequality, employment growth, and functional distribution of income. 

\section{Data}\label{sec:data}

\paragraph{EUROSTAT's Structural Business Survey.}\label{sec:sbs}

The main data source used in this paper is the Eurostat Structural Business Statistics (SBS). The SBS describe the structure and performance of industrial sectors in the in European countries. They cover the ‘business economy’ (NACE Rev. 2 sections B to N and division 95) which includes: i) Industry; ii) Construction; iii) Distributive trades; and iv) Services. The data on industries is aggregated at the 4 digit level of the NACE classification, for a total of 218 NACE industries. SBS are based upon data for enterprises, which is obtained for national statistical offices and aggregated at the industry level based on the enterprise’s principal activity. Businesses in financial services (NACE Rev. 2 Section K) are excluded because of their specific characteristics and the unavailability of information in this industry, often due to confidentiality issues. In addition, SBS does not cover agriculture, forestry and fishing (NACE A), public administration (NACE O) and non-market services such as education (NACE P) and health (NACE Q).

As the analysis conducted here relies on patterns of industrial employment, we use the SBS data to describe the employment share of 4 digit NACE industries. To do so, we extract the SBS indicator that quantifies the number of persons employed\footnote{Following the Eurostat's SBS definition, this indicator "is defined as the total number of persons who work in the observation unit (inclusive of working proprietors, partners working regularly in the unit and unpaid family workers), as well as persons who work outside the unit who belong to it and are paid by it (e.g. sales representatives, delivery personnel, repair and maintenance teams). It excludes manpower supplied to the unit by other enterprises, persons carrying out repair and maintenance work in the enquiry unit on behalf of other enterprises, as well as those on compulsory military service." Source: \url{https://ec.europa.eu/eurostat/cache/metadata/en/sbs_esms.htm\#meta_update1663843213928}.} in each industry. In order to maximise data availability, we restrict the analysis to the 27 countries that were part of the European Union until 2018, observed between 2010 and 2018. 

One limitation of the SBS data is that it presents a large number of missing data across countries, industries and years. Given that the the construction of the indicators used in the analysis (such as the labour-weighted fitness illustrated in section \ref{sec:lwf}) requires full series of industrial employment shares across countries and years, we have resorted to a data reconstruction strategy, based on linear interpolation and extrapolation at constant values. This strategy has proved to be the best performing in terms of its ability to correctly predict missing values. A detailed description of the data reconstruction strategy, and a comparison between competing methods, is presented in Appendix \ref{sec:data-rec}.

\paragraph{ARDECO.}\label{sec:ardeco}

The Annual Regional Database of the European Commission's Directorate General for Regional and Urban Policy (ARDECO),\footnote{The ARDECO dataset is available, free of charge, from the European Commission's website: \url{https://knowledge4policy.ec.europa.eu/territorial/ardeco-database_en}.} is a dataset curated and maintained by the Commission's Joint Research Centre. It includes information on European countries and regions at different levels of disaggregation. While the dataset is based on the Eurostat's SBS illustrated in the previous section, the data is complemented -- where needed -- with national and international sources. The data covers different topics, such as population, employment, labour costs, domestic product and capital formation. For the purpose of the analysis carried out in this paper, indicators on population, GDP per capita (constant prices), employment growth, compensation of employees (at constant prices) and gross value added (at constant prices) have been selected. The last two indicators are used to construct a measure of labour share of the economy, computed as the ratio of employee's aggregate compensation over gross value added, in each country and year.

\paragraph{ILO Statistics on Labour Income and Inequality.}\label{sec:ILO}

The data on wage inequality is obtained from the International Labour Organisation's Statistics on Labour Income and Inequality. We focus on data regarding the labour income distribution,\footnote{Available at: \url{https://www.ilo.org/shinyapps/bulkexplorer46/?id=LAP_2LID_QTL_RT_A}.} which despite being based on modelled estimates, it provides information on average wages at all deciles of the labour income distribution across the countries in the sample, for the entire time period covered (2010 -- 2018). In particular, we use information on labour income at the ninth, fifth and first deciles to produce wage ratios between ninth and first, fifth and first, and ninth and fifth deciles to capture the degree of inequality in labour income at different points of the distribution. 

\paragraph{World Development Indicators.}\label{sec:WDI}

The data described above is complemented with additional information extracted from the World Bank's World Development Indicators to further characterise the profile of the countries in our sample, and to mitigate the omitted variable bias. As we are focusing on time-varying characteristics of countries that have somehow to do with their productive structure and technological capabilities, we use information on the share of exports and share of R\&D investments over total GDP.

\section{Methods}\label{sec:methods}

\subsection{From Inferred Comparative Advantages to a measure of Labour-weighted Fitness}\label{sec:biwcm}

A common starting point in the construction of Economic Complexity indicators is the observation of countries' specialisation profiles across sectors or products. In practice, these profiles are typically identified through measures of comparative advantage, most notably Balassa's (\citeyear{Balassa1965}) Revealed Comparative Advantage -- which defines specialisation of country $c$ in industry $i$ by comparing the observed relative weight of an activity in a given country to its expected value -- that is, global average share of the typical country in the same activity. Specialisation is therefore defined by this share being larger than one. Nevertheless, this approach has a number of shortcomings: as pointed out by \citep{bruno_inferring_2023}, RCA is sensitive to extreme values and skewed distributions -- like it is the case for industrial employment data -- with outliers weighting disproportionately on the expected value used as denominator in the RCA formula. Moreover, RCA treats all country-activity pairs as independent, missing correlations that naturally arise in real networks. 

In order to address these shortcomings,\footnote{The issues noted above and described in \citet{bruno_inferring_2023} affect skewed data -- like industrial employment data -- to a larger extent as compared to activity data traditionally used in Economic Complexity, such as international trade data.} in this paper we adopt a recent approach proposed by \cite{bruno_inferring_2023}, the Inferred Comparative Advantage (ICA), to construct the binary country-sector matrix $M_{c,i}$.  Rather than relying on 'raw' RCA thresholds, ICA compares the observed allocation of labour across country-sector pairs with the distribution implied by an entropy-based unbiased null-model -- the  Bipartite Weighted Configuration Model (BiWCM) -- that preserves two key aggregate features of the matrix, namely total employment at the country level and total employment at the sector level across all analysed countries. Further details on the methodology are provided in Appendix \ref{sec:ICA}. 

Incorporating a null model allows us to identify specialisation patterns in an unbiased and more statistically robust way. For instance, Figure \ref{fig:biwcm} reports the p-values associated to the empirical values of $M_{c,i}$ using a colour scale ranging from blue (zero), representing non significant country-sector links, to red (one), representing instead statistically significant links. Before feeding the ICA ($M_{c,i}$) matrix described in Figure \ref{fig:biwcm} into the EFC algorithm outlined in Equation \eqref{eq:fitness}, we implement an additional step to ensure comparability of Fitness and Complexity values across years. 

Since the EFC algorithm is computed separately for each year, the numerical values of Fitness depend on the structure of the underlying country--sector matrix and may therefore vary in scale over time. Raw Fitness values are thus not directly comparable across years. Following \citep{mazzilli2024equivalence}, we address this issue by adding a dummy country specialised in all sectors and rescaling all Fitness values relative to this fixed benchmark, thereby ensuring temporal comparability. More details on this procedure are provided in Appendix \ref{sec:fitscale}.

\subsection{Econometric framework}\label{sec:model}
It is now possible to compute the Industrial Complexity Index $Q_i$ used to construct our measure of Labour-Weighted Fitness, as described in Equation \ref{eq:lwf}, which in turn is decomposed into a within and a between term (Equation \ref{eq:dec4}). In particular, as mentioned in Section \ref{sec:lwf}, we are interested in the decomposition term that captures the shift of labour shares across industries, that is, the between term, which captures the process of structural change isolated from the total change in labour-weighted Fitness. In other words, the between component of our decomposition pins down the changes in total fitness that derive from labour moving towards more complex industries, rather than industries becoming more complex themselves. We use this term to explore its correlation with employment growth, wage inequality and the labour share of the economy, using the following specification: 

\begin{equation}
    \label{eq:model}
    Y_{c,t} = \alpha + \beta_1 between_{c,t} + \beta_2 within_{c,t} + \beta_3 \Delta LWF_{c,t} + \beta_4 X_{c,t} + \sigma_c + \tau_t + \epsilon_{c,t}
\end{equation}

\noindent where $between_{c,t}$, $within_{c,t}$ and $\Delta LWF_{c,t}$ are respectively the between term, within term, and sum of the previous two, as described by Equation  \ref{eq:dec4}; in all cases, these quantities are computed for $k = 1$. 
$Y_{c,t}$ is a placeholder for measures of employment growth, wage inequality and labour share of the economy. Employment growth is computed as the share of employed, working age population in each country at time $t$ over the same share of employment in the previous time period, as obtained from the ARDECO dataset: 

\begin{equation}
    \label{eq:empgrowth}
    g_{c,t} = \frac{emp_{c,t}}{emp_{c,t-1}}.
\end{equation}

\noindent Wage inequality is constructed using the ILO Statistics on Labour Income and Inequality described in Section \ref{sec:ILO}. In order to capture the effect of structural change on the distribution of income, we take three different measures of wage inequality: 9$^{th}$ to 1$^{st}$ percentile, 9$^{th}$ to 5$^{th}$ percentile, and 5$^{th}$ to 1$^{st}$ percentile: 

\begin{equation}
    \label{eq:ineq}
    ineq_{c,t}^{p^{th}/q^{th}} = \frac{wage_{c,t}^{pth}}{{wage_{c,t}^{qth}}}
\end{equation}

\noindent with the superscripts $p$ and $q$ indicating either the 9$^{th}$, 5$^{th}$ or 1$^{st}$ percentile. 
Finally, we use the ARDECO data to compute a measure of the labour share of the economy in each country and year, taking the ratio of the total wages paid to workers over the total value added, all measured at constant prices (2015): 

\begin{equation}
    \label{eq:labshare}
    labshare_{c,t} = \frac{wage_{c,t}}{VA_{c,t}}
\end{equation}

\noindent Finally, $X_{c,t}$ is a vector of time varying controls at the country level, including GDP per capita, population (from ARDECO), share of R$\&$D investments over GDP, share of exports over GDP (from the World Development Indicators). All equations are estimated using OLS country ($\sigma_c$) and time ($\tau_t$) fixed effects, with errors clustered by country and year. 

\section{Results}\label{sec:results}

\subsection{Descriptive results}\label{sec:desc}

\begin{figure}[!ht]
    \centering
    
    \caption{ICA Specialisation Matrix}
    \includegraphics[width=0.8\linewidth]{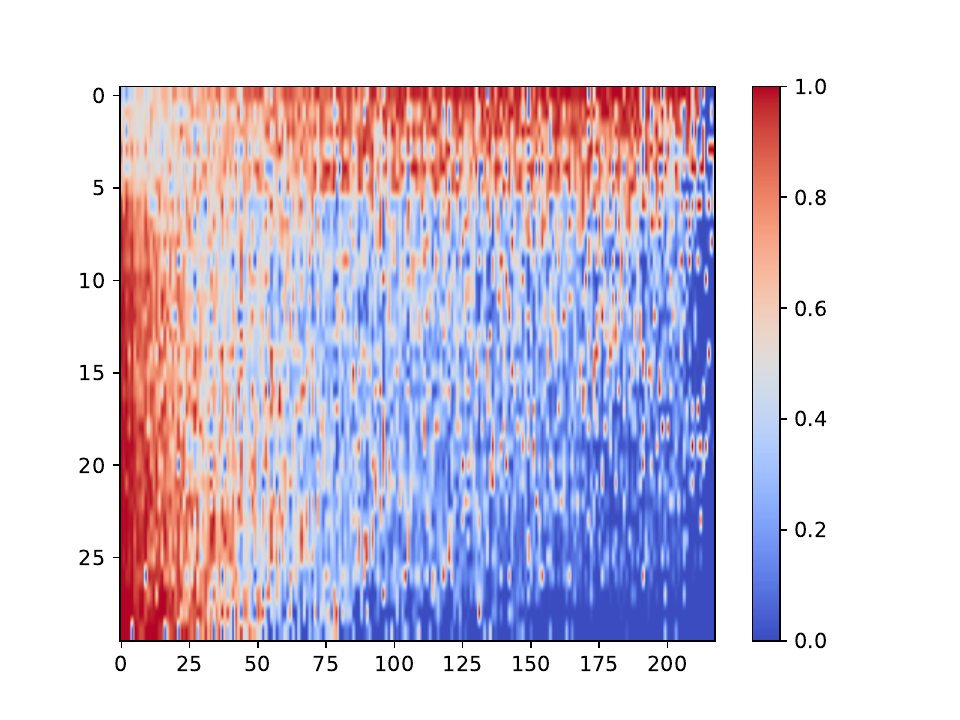}
    \begin{threeparttable}
        \begin{tablenotes}[flushleft]
        \footnotesize
        \item Notes: The figure reports the ICA specialisation matrix based on the BiWCM filtering procedure, evaluated for year 2018, where countries are displayed in the y-axis (higher fitness on the top) and the sectors in which they specialised in the x-axis (more complex industries to the right). Coloured cells indicate the statistical significance attributed to specialisation of a country in an industry. The colours on the colour bar report the (1 - pvalue) figures, with higher (statistically significant) values in red, and lower (non statistically significant) values in blue. 
    \end{tablenotes}

\end{threeparttable}

    \label{fig:biwcm}
\end{figure}

We begin by commenting on the patterns revealed by the specialisation matrix, which has been obtained by filtering the country-industry employment matrix for the year 2010 as described in Section \ref{sec:biwcm}. Figure \ref{fig:biwcm} contains information on the statistical significance of individual country-industry linkages, with values going from more significant (red) to less significant (blue). The rows of the matrix represent countries, while the columns identify industries. Countries have been ordered from higher Fitness (top) to lower Fitness (bottom), and industries are sorted from the least (left) to the most (right) complex. 

\bigskip

Rather, it points to a highly ordered structure in which diversification and specialization are systematically related. In economic terms, industries observed in low-Fitness countries are generally also observed in high-Fitness countries, whereas high-Fitness countries are distinguished by their ability to operate in many additional, less ubiquitous activities. At a first visual inspection, the matrix displays a nested structure \citep{Bustos2012} -- a typical feature of specialisation matrices \citep{Patelli2023} indicating the presence of a hierarchical structure in the observed activities.  Intuitively, nestedness means that the specialisation pattern of less diversified countries tends to form a subset of that of more diversified ones: industries in which less diversified economies are significantly present are typically also present in more diversified ones, while the latter are active in a wider range of industries. In the matrix, this appears as a roughly triangular pattern in the distribution of the statistically significant links between countries and industries. This type of structure, originally studied in evolutionary biology, has become central in the economic complexity literature \citep{neffke2026economic}. 

\begin{figure}[!ht]
    \centering
    \caption{Fitness rankings (2010-2018).}
    \includegraphics[width=0.8\linewidth]{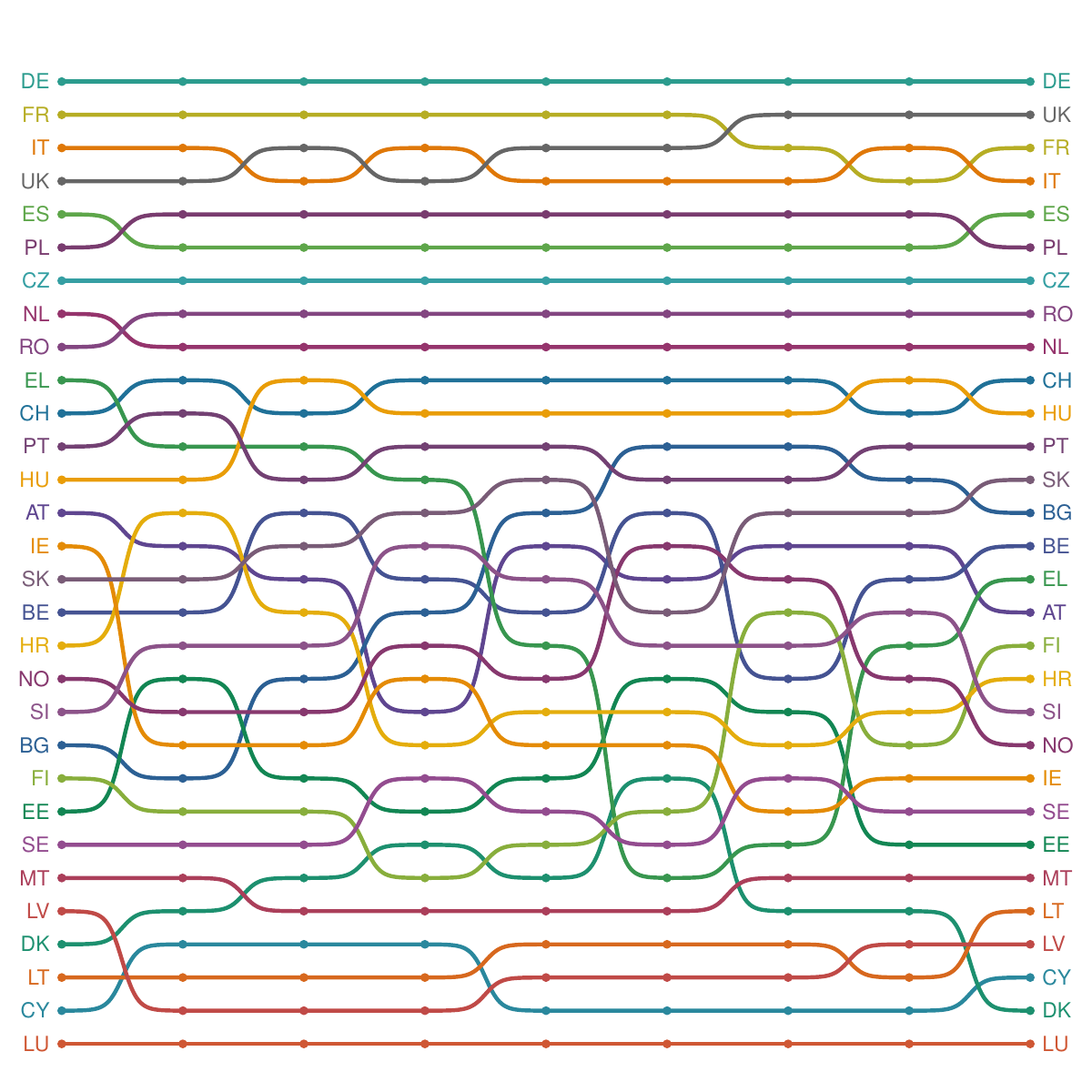}
    \begin{threeparttable}
        \begin{tablenotes}[flushleft]
        \footnotesize
        \item \textit{Notes}: The figure displays the evolution of Fitness country rankings based on ICA specialisation matrices between years 2010 and 2018. Countries are sorted in descending order by their Fitness in 2010 (left axis) and 2018 (right axis). Dots along the lines in the spaghetti plot identify years between 2010 and 2018. 
    \end{tablenotes}

\end{threeparttable}
    \label{fig:rank}
\end{figure}

In the case described in Figure \ref{fig:biwcm}, the economic interpretation of nestedness holds only in part; the top left corner -- where high-Fitness countries and low-Complexity industries are located -- appears to be less populated than expected in a fully nested matrix. While a similar pattern emerges also in other systems,  such as in the specialisation matrix describing the scientific production of countries \citep{Patelli2023}, in this case we find the observed result consistent with the process of structural change. High-Fitness countries tend to have higher diversification and develop a comparative advantage in capital- or knowledge-intensive industries, at the expense of more labour-intensive (and low-complexity) ones. As a result, countries with  comparative advantages in labour intensive activities (often reflecting lower labour costs) tend to specialise in those industries and exhibit lower levels of diversification. In broad terms, the sample of European countries we analyse appears to be split into two main groups at different stages of structural transformation: a smaller set of more mature, service-based economies and a larger group of less diversified countries. The evolution of Fitness rankings reported in Figure  \ref{fig:rank} clearly illustrates this divide.  The upper part of the ranking is stably occupied by the more mature Western European economies -- most notably Germany, France, Italy, United Kingdom and Spain -- followed by a second group formed mainly by emerging Central and Eastern European (CEE) countries. 

\begin{figure}[!ht]
    \centering
    \caption{Labour-weighted Fitness decomposition (2010-2018).}
    \includegraphics[width=0.8\linewidth]{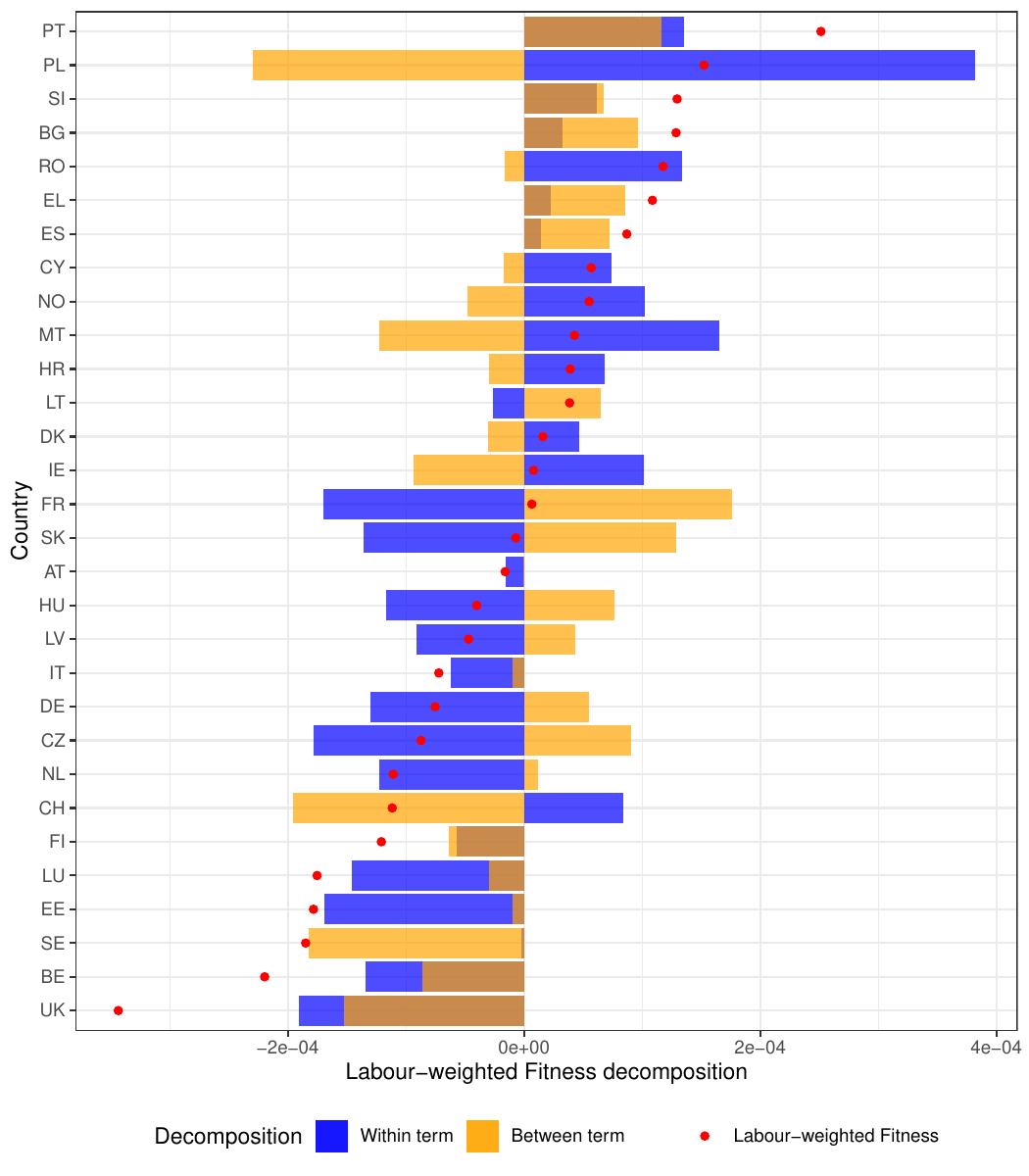}
    \begin{threeparttable}
        \begin{tablenotes}[flushleft]
        \footnotesize
        \item \textit{Notes}: For each country along the vertical axis, the plot reports the changes in Labour-weighted Fitness between 2010 and 2018 (red dot) and its  between and within-components, respectively represented by the yellow and blue bars. Countries are sorted in descending order by magnitude of change in Labour-weighted Fitness. 
    \end{tablenotes}
\end{threeparttable}
    \label{fig:dec_t0}
\end{figure}

We now move towards an analysis of the variation in LWF over the time period 2010--2018. Figure \ref{fig:dec_t0} shows the difference in LWF between 2010 and 2018, computed as per Equation \ref{eq:dec4}, with its respective between and within-component. We shall recall that the former measures the contribution of structural change -- defined here as the movement of labour to more complex industries -- to the variation in LWF over the time period under consideration; while the latter picks up the change in LWF due to increasing complexity of the industries in which countries develop a (inferred) comparative advantage. The barplot in Figure \ref{fig:dec_t0} indicates that the between- and within-component are often anticorrelated. This is consistent with the reasonable assumption that as industries become more complex, they also become less labour intensive. To make an example, the industries that exhibits the largest increase in complexity is Mining of Hard Coal -- an industry that has become progressively more technologically intensive, and less able to absorb workers \citep{paredes2021automation}. 
The within-component captures if the industries in which a country is specialised become more complex -- for instance, as a result of technological upgrading within the industry. By doing so, labour is shed by such complex industries, flowing into industries with lower complexity. This employment reallocation is thus often associated with a between-component of the opposite sign to the within-component. 

\begin{figure}[!ht]
    \centering
    \caption{Labour-weighted Fitness decomposition (2010-2014 and 2014-2018).}
    \includegraphics[width=1\linewidth]{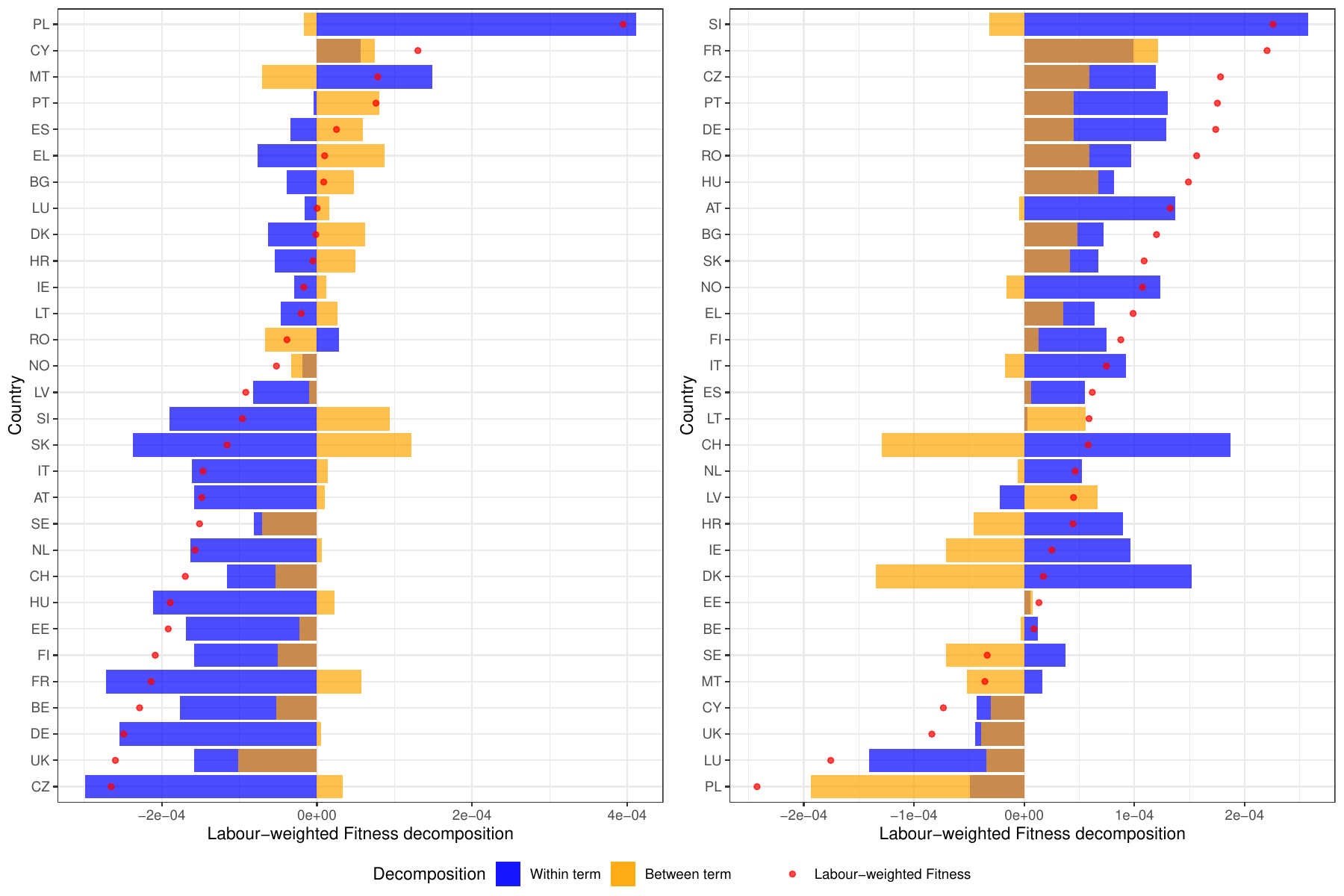}
        \begin{threeparttable}
        \begin{tablenotes}[flushleft]
        \footnotesize
        \item \textit{Notes}: For each country along the vertical axis, the plot reports the changes in Labour-weighted Fitness (red dot) and its  between- and within-component, respectively represented by the yellow and blue bars. The left panel refers to the period between 2010 and 2014, while the right panel refers to the period between 2014 and 2018. Countries are sorted in descending order by magnitude of change in Labour-weighted Fitness. 
    \end{tablenotes}
\end{threeparttable}
    
    \label{fig:dec_t1}
\end{figure}

To have a clearer idea of what happened in European countries over the 2010--2018 period, Figure \ref{fig:dec_t1} describes the same quantities as Figure \ref{fig:dec_t0}, but for two different time periods: 2010--2014 (left panel) and 2014--2018 (right panel). The figure indicates that in the first time period the variation in LWF fitness was negative for most countries, mostly driven by the decreasing within industry complexity (blue bars). It must be noted that this period coincides with the aftermaths of the 2008 financial crises, which has severely compromised economic activity in Europe across the board. In the following period, however, European countries show a much better performance, with most countries increasing their LWF, in many case due to within industry complexity and between industry reallocation of labour. 

Poland represents a particularly interesting and illustrative case, which is an outlier in both periods. In 2010--2014, the country shows a very substantial within-industry complexity growth, which  accounts for almost the entirety of the increase in labour-weighted Fitness. In the subsequent time period, instead, the within-component becomes negligible, while the between-component turns negative. This pattern points to a two-stage process: an initial phase of upgrading, in which complexity increased within existing sectors with little accompanying reallocation of labour across industries, followed by a phase in which these same activities displayed weaker labour absorption, leading employment to shift towards less complex sectors. In this sense, the negative between component may reflect the declining labour intensity of a productive structure increasingly oriented towards more complex and knowledge-intensive activities. This interpretation is consistent both with the earlier literature on FDI-led post-socialist restructuring, which emphasises the role of foreign-owned manufacturing plants in technological upgrading and structural transformation \citep{domanski2003industrial}, and with more recent evidence documenting Poland’s capability accumulation, increasingly diversified productive structure, deeper integration into global value chains, and the growing importance of knowledge-intensive services in the national economy \citep{OECD2025PolandFDI}.

\subsection{Structural change and employment growth, wage inequality, and functional distribution of income}\label{sec:results_regs}

In this section we report the results of the estimation of Equation \ref{eq:model} using the different outcome variables for employment growth, wage inequality and functional distribution of income. It should borne in mind that the results presented below aim exclusively at describing the correlation between structural change and the outcome variables, without any implication for their causal relationships. 

Table \ref{tab:reg_empgrowth} summarises the relationship between structural change (the variable \textit{Between} in the table) and employment growth. Looking at columns (1) and (4), the between-component of the LWF structural decomposition shows a negative correlation with employment growth at the country level. This indicates that when countries experience a labour flow from low- to high-complexity industries, they also experience slower rates of employment growth. This finding seems to corroborate the hypothesis that high-complexity industries contribute less to aggregate employment growth, due to their higher levels of knowledge intensity, and lower labour intensity, as they require fewer (and likely more skilled) workers. It is worth mentioning that neither the within-component of the variation in LWF (column 2), nor the sum of both components (column 3) shows any significant correlation with employment growth, highlighting the importance of structural change in explaining outcomes related to the creation of employment. The coefficient on the between-component remains significant also when the within-component is also included in the regression. 

\begin{table}[!ht]
   \caption{\label{tab:reg_empgrowth} Employment growth regressions}
   \centering
   \footnotesize

   \begin{threeparttable}

   \begin{tabular}{lcccc}
      \midrule \midrule
      Dependent Variable: & \multicolumn{4}{c}{Yearly employment rate growth ($\%$)}\\
       & \multicolumn{4}{c}{FE OLS} \\ 
      Model:              & (1) & (2) & (3) & (4)\\  
      \midrule
      \emph{Variables}\\
      Between             & -7,313.359$^{**}$ &               &              & -7,527.217$^{**}$\\   
                          & (3,229.347)       &               &              & (3,248.861)\\   
      Within              &                   & 1,266.110     &              & 1,412.623\\   
                          &                   & (1,121.540)   &              & (1,123.097)\\   
      $\Delta LWF$          &                   &               & 279.081      &   \\   
                          &                   &               & (993.907)    &   \\ 
      Population (log)    & -19.942           & -19.817       & -19.927      & -19.785\\   
                          & (12.878)          & (13.098)      & (13.057)     & (12.968)\\   
      GDPpc (log)         & -6.179$^{*}$      & -5.824        & -6.046       & -5.859\\   
                          & (3.563)           & (3.572)       & (3.556)      & (3.570)\\   
      R$\&$D ($\%$ GDP)   & 0.169$^{**}$      & 0.175$^{***}$ & 0.174$^{**}$ & 0.170$^{**}$\\   
                          & (0.064)           & (0.063)       & (0.063)      & (0.063)\\   
      Exports ($\%$ GDP)  & 0.142             & 0.232         & 0.177        & 0.222\\   
                          & (0.761)           & (0.736)       & (0.756)      & (0.718)\\   
  
      \midrule
      \emph{Fixed-effects}\\
      Country             & Yes & Yes & Yes & Yes\\  
      Year                & Yes & Yes & Yes & Yes\\  
      \midrule
      \emph{Fit statistics}\\
      Observations        & 224   & 224   & 224   & 224\\  
      R$^2$               & 0.526 & 0.522 & 0.520 & 0.528\\  
      Within R$^2$        & 0.233 & 0.225 & 0.223 & 0.236\\  
      \midrule \midrule
   \end{tabular}

   \begin{tablenotes}[flushleft]
   \footnotesize
   \item \textit{Notes:} Clustered (Country-Year) standard errors reported in parentheses. 
   All specifications include country and year fixed effects. 
   The dependent variable is the growth rate in employment between year $t$ and $t-1$ . 
   \textit{Between} captures the between-component of the yearly change in Labour Weighted Fitness, while 
   \textit{Within} captures the within-component of the Labour Weighted Fitness decomposition (both from Equation \ref{eq:dec4}). $\Delta LWF$ denotes the change in the Labour Weighted Fitness index. 
   \textit{GDPpc (log)} is log GDP per capita in US dollars (constant 2015); \textit{Population (log)} is log total population; 
  \textit{ R\&D ($\%$ GDP)} is gross domestic expenditure on research and development as a percentage of GDP; 
   \textit{Exports ($\%$ GDP)} is total exports as a percentage of GDP. 
   $^{***}p<0.01$, $^{**}p<0.05$, $^{*}p<0.1$.
   \end{tablenotes}

   \end{threeparttable}
\end{table}

We now move to the relationship between structural change and wage inequality. First, we look at the link between structural change and the top-to-bottom inequality, i.e. the wage gap between top (9$^{th}$ decile) and bottom (1$^{st}$ decile) wages. As can be appreciated by the negative and significant coefficients of the between-component (alone and together with the within-component respectively in column 1 and column 4) in Table \ref{tab:reg_wage91}, the 9$^{th}$/1$^{st}$ wage ratio appears to contract as a result of labour reallocation towards knowledge-intensive industries. However, it is not clear whether this result is driven by lower wages in the top of the distribution, or higher wages at the bottom. To further explore this, Tables \ref{tab:reg_wage51} and \ref{tab:reg_wage95} show the results for upper- and lower-tail inequality,  the 5$^{th}$/1$^{st}$ and 9$^{th}$/5$^{th}$ wage ratios, respectively. Interestingly, structural change is associated with a narrowing of the wage gap between median and bottom wages (columns 1 and 4 in Table  \ref{tab:reg_wage51}), whereas it does not appear to be significantly associated with the wage gap between top and median wages (columns 1 and 4 in Table  \ref{tab:reg_wage95}).

The combination of these results, therefore, suggests that the reduction in the wage gap between top and bottom wages is driven mainly by changes at the bottom of the wage distribution. In order to rationalise this result, it is worth going back to the specialisation matrix described in Figure \ref{fig:biwcm}. As countries move to more complex industries, they drop the least complex ones, as indicated by the "emptier" top-left corner of the matrix. Assuming a positive correlation between complexity and average wages, it can be inferred that the reduction in wage inequality is driven by the fact that the very low-paid jobs in the least complex sectors disappear, making the 1$^{st}$ decile of the salary distribution go up. 

\begin{table}[!ht]
   \caption{\label{tab:reg_wage91} Wage inequality regressions: ratio $9^{th} / 1^{st}$ dec.}
   \centering
   \footnotesize

   \begin{threeparttable}

   \begin{tabular}{lcccc}
      \midrule \midrule
      Dependent Variable: & \multicolumn{4}{c}{Ratio $9^{th} / 1^{st}$ dec.}\\
       & \multicolumn{4}{c}{FE OLS} \\ 
      Model:              & (1) & (2) & (3) & (4)\\  
      \midrule
      \emph{Variables}\\
      Between             & -8,201.860$^{***}$ &           &            & -8,142.939$^{***}$\\   
                          & (2,884.048)        &           &            & (2,893.616)\\   
      Within              &                    & -547.693  &            & -389.195\\   
                          &                    & (671.490) &            & (744.401)\\ 
      $\Delta LWF$          &                    &           & -1,372.345 &   \\   
                          &                    &           & (956.596)  &   \\   
      GDPpc (log)         & -3.781             & -3.831    & -4.031     & -3.870\\   
                          & (2.770)            & (2.863)   & (2.861)    & (2.778)\\   
      Population (log)    & -5.596             & -5.675    & -5.763     & -5.640\\   
                          & (7.401)            & (7.552)   & (7.514)    & (7.424)\\   
      R$\&$D ($\%$ GDP)   & -0.013             & -0.008    & -0.010     & -0.014\\   
                          & (0.024)            & (0.025)   & (0.024)    & (0.024)\\   
      Exports ($\%$ GDP)  & -1.080             & -1.091    & -1.141     & -1.102\\   
                          & (0.995)            & (1.000)   & (1.000)    & (1.002)\\   
      \midrule
      \emph{Fixed-effects}\\
      Country             & Yes & Yes & Yes & Yes\\  
      Year                & Yes & Yes & Yes & Yes\\  
      \midrule
      \emph{Fit statistics}\\
      Observations        & 224   & 224   & 224   & 224\\  
      R$^2$               & 0.906 & 0.903 & 0.904 & 0.906\\  
      Within R$^2$        & 0.073 & 0.044 & 0.051 & 0.074\\  
      \midrule \midrule
   \end{tabular}

   \begin{tablenotes}[flushleft]
   \footnotesize
   \item \textit{Notes:} Clustered (Country-Year) standard errors reported in parentheses. 
   All specifications include country and year fixed effects. 
   The dependent variable is the wage ratio between the 9$^{th}$ and 1$^{st}$ deciles of the wage distribution. 
   \textit{Between} captures the between-component of the yearly change in Labour Weighted Fitness, while 
   \textit{Within} captures the within-component of the Labour Weighted Fitness decomposition (both from Equation \ref{eq:dec4}). $\Delta LWF$ denotes the change in the Labour Weighted Fitness index. 
   \textit{GDPpc (log)} is log GDP per capita in US dollars (constant 2015); \textit{Population (log)} is log total population; 
   \textit{R\&D ($\%$ GDP)} is gross domestic expenditure on research and development as a percentage of GDP; 
   \textit{Exports ($\%$ GDP)} is total exports as a percentage of GDP. 
   $^{***}p<0.01$, $^{**}p<0.05$, $^{*}p<0.1$.
   \end{tablenotes}

   \end{threeparttable}
\end{table}

\begin{table}[!ht]
   \caption{\label{tab:reg_wage51} Wage inequality regressions, wages ratio ($5^{th} / 1^{st}$ dec.)}
   \centering
   \footnotesize
   \begin{threeparttable}
       
    \begin{tabular}{lcccc}
      \tabularnewline \midrule \midrule
      Dependent Variable: & \multicolumn{4}{c}{Ratio $5^{th} / 1^{st}$ dec.}\\
       & \multicolumn{4}{c}{FE OLS} \\ 
      Model:              & (1)                & (2)       & (3)       & (4)\\  
      \midrule
      \emph{Variables}\\
      Between             & -4,368.036$^{***}$ &           &           & -4,358.609$^{***}$\\   
                          & (1,552.729)        &           &           & (1,555.969)\\   
      Within              &                    & -147.110  &           & -62.273\\   
                          &                    & (275.997) &           & (290.821)\\   
      $\Delta LWF$          &                    &           & -607.034  &   \\   
                          &                    &           & (405.836) &   \\   \textbf{}
      GDPpc (log)         & -1.009             & -1.003    & -1.113    & -1.023\\   
                          & (1.122)            & (1.173)   & (1.176)   & (1.126)\\   
      Population (log)    & -4.232             & -4.258    & -4.308    & -4.239\\   
                          & (3.129)            & (3.226)   & (3.211)   & (3.144)\\   
      R$\&$D ($\%$ GDP)   & -0.007             & -0.005    & -0.005    & -0.008\\   
                          & (0.011)            & (0.011)   & (0.011)   & (0.011)\\   
      Exports ($\%$ GDP)  & -0.414             & -0.412    & -0.440    & -0.418\\   
                          & (0.413)            & (0.420)   & (0.420)   & (0.418)\\   
      
      \midrule
      \emph{Fixed-effects}\\
      Country             & Yes                & Yes       & Yes       & Yes\\  
      Year                & Yes                & Yes       & Yes       & Yes\\  
      \midrule
      \emph{Fit statistics}\\
      Observations        & 224                & 224       & 224       & 224\\  
      R$^2$               & 0.907              & 0.903     & 0.903     & 0.907\\  
      Within R$^2$        & 0.090              & 0.048     & 0.055     & 0.090\\  
      \midrule \midrule

   \end{tabular}
     \begin{tablenotes}[flushleft]
   \footnotesize
   \item \textit{Notes:} Clustered (Country-Year) standard errors reported in parentheses. 
   All specifications include country and year fixed effects. 
   The dependent variable is the wage ratio between the 5$^{th}$ and 1$^{st}$ deciles of the wage distribution. 
   \textit{Between} captures the between-component of the yearly change in Labour Weighted Fitness, while 
   \textit{Within} captures the within-component of the Labour Weighted Fitness decomposition (both from Equation \ref{eq:dec4}). $\Delta LWF$ denotes the change in the Labour Weighted Fitness index. 
   \textit{GDPpc (log)} is log GDP per capita in US dollars (constant 2015); \textit{Population (log)} is log total population; 
   \textit{R\&D ($\%$ GDP)} is gross domestic expenditure on research and development as a percentage of GDP; 
   \textit{Exports ($\%$ GDP)} is total exports as a percentage of GDP. 
   $^{***}p<0.01$, $^{**}p<0.05$, $^{*}p<0.1$.
   \end{tablenotes}

   \end{threeparttable}
\end{table}

\begin{table}[!ht]
   \caption{\label{tab:reg_wage95} Wage inequality regressions, wages ratio ($9^{th} / 5^{st}$ dec.)}
   \centering
   \footnotesize
   \begin{threeparttable}
      \begin{tabular}{lcccc}
      \tabularnewline \midrule \midrule
      Dependent Variable: & \multicolumn{4}{c}{Ratio $9^{th} / 5^{th}$ dec.}\\
       & \multicolumn{4}{c}{FE OLS} \\ 
      Model:              & (1)           & (2)           & (3)           & (4)\\  
      \midrule
      \emph{Variables}\\
      Between             & -44.173       &               &               & -39.238\\   
                          & (149.532)     &               &               & (147.483)\\   
      Within              &               & -33.360       &               & -32.596\\   
                          &               & (70.242)      &               & (71.158)\\   
      $\Delta LWF$          &               &               & -33.438       &   \\   
                          &               &               & (64.304)      &   \\   
      Population (log)    & 0.556         & 0.553         & 0.553         & 0.553\\   
                          & (0.502)       & (0.498)       & (0.498)       & (0.499)\\   
      GDPpc (log)         & -0.346$^{**}$ & -0.353$^{**}$ & -0.353$^{**}$ & -0.353$^{**}$\\   
                          & (0.168)       & (0.162)       & (0.163)       & (0.162)\\   
      R$\&$D ($\%$ GDP)   & 0.001         & 0.001         & 0.001         & 0.001\\   
                          & (0.002)       & (0.002)       & (0.002)       & (0.002)\\   
      Exports ($\%$ GDP)  & -0.035        & -0.037        & -0.037        & -0.037\\   
                          & (0.050)       & (0.048)       & (0.049)       & (0.049)\\   
      
      \midrule
      \emph{Fixed-effects}\\
      Country             & Yes           & Yes           & Yes           & Yes\\  
      Year                & Yes           & Yes           & Yes           & Yes\\  
      \midrule
      \emph{Fit statistics}\\
      Observations        & 224           & 224           & 224           & 224\\  
      R$^2$               & 0.939         & 0.939         & 0.939         & 0.939\\  
      Within R$^2$        & 0.109         & 0.110         & 0.110         & 0.110\\  
      \midrule \midrule
   \end{tabular}
      \begin{tablenotes}[flushleft]
   \footnotesize
   \item \textit{Notes:} Clustered (Country-Year) standard errors reported in parentheses. 
   All specifications include country and year fixed effects. 
   The dependent variable is the wage ratio between the 9$^{th}$ and 5$^{th}$ deciles of the wage distribution. 
   \textit{Between} captures the between-component of the yearly change in Labour Weighted Fitness, while 
   \textit{Within} captures the within-component of the Labour Weighted Fitness decomposition (both from Equation \ref{eq:dec4}). $\Delta LWF$ denotes the change in the Labour Weighted Fitness index. 
   \textit{GDPpc (log)} is log GDP per capita in US dollars (constant 2015); \textit{Population (log)} is log total population; 
   \textit{R\&D ($\%$ GDP)} is gross domestic expenditure on research and development as a percentage of GDP; 
   \textit{Exports ($\%$ GDP)} is total exports as a percentage of GDP. 
   $^{***}p<0.01$, $^{**}p<0.05$, $^{*}p<0.1$.
   \end{tablenotes}

   \end{threeparttable}
\end{table}

Finally, Table \ref{tab:reg_labshare} displays the results of regressing the labour share (defined as the ratio between aggregate wages and value added in the economy) on the between-component of LWF. Columns (1) and (4) show a positive relationship between the two, indicating that, as workers move to more complex industries, the labour share tends to increase. This indicates that structural change towards more complex industries is associated with a higher share of labour in total value added. To interpret this finding, we should rely once more on the evidence that high-complexity industries pay, on average, higher wages, as shown by \cite{Sbardella2017, aufiero2024mapping}. Bearing in mind the diversification patterns displayed by Figure \ref{fig:biwcm}, coupled with the results on employment growth in Table \ref{tab:reg_empgrowth}, if employment growth slows down as a result of structural change towards knowledge-intensive industries, the increase in the labour share of the economy is due to be driven by higher average wages rather than broader employment expansion. This positive labour share-structural change association is particularly noteworthy because it contrasts with the broad declining trend in the labour share documented for European economies over recent decades \citep{Arpaia2009,stockhammer2012have, Dimova2019}. While at the aggregate level the literature has generally associated this decline with capital deepening, technological change, globalisation, and weakening labour market institutions, our findings suggest that, holding other factors constant, structural upgrading towards more complex sectors appears as a countervailing force to the aggregate downward trend, suggesting that this specific component of structural transformation may partly offset the broader pressures that have contributed to the decline in the labour share. This positive effect is also consistent with what observed by \citet{hubmer2026}, who document a rise in the labour share of typical firms, showing how larger firms are the ones that drive the capital-labour substitution driven by automation. An expansion of industries dominated by firms similar to the typical firm would be associated also to a growth in the labour share. 

\begin{table}[!ht]
   \caption{\label{tab:reg_labshare} Labour share regressions}
   \centering
   \footnotesize
   \begin{threeparttable}
   
\begin{tabular}{lcccc}
      \tabularnewline \midrule \midrule
      Dependent Variable: & \multicolumn{4}{c}{Labour share (constant 2015 USD)}\\
       & \multicolumn{4}{c}{FE OLS} \\ 
      Model:              & (1)               & (2)           & (3)           & (4)\\  
      \midrule
      \emph{Variables}\\
      Between             & 10,825.259$^{**}$ &               &               & 11,050.893$^{**}$\\   
                          & (4,288.268)       &               &               & (4,255.691)\\   
      Within              &                   & -1,275.304    &               & -1,490.403\\   
                          &                   & (1,437.922)   &               & (1,428.899)\\   
      $\Delta LWF$          &                   &               & 99.792        &   \\   
                          &                   &               & (1,324.813)   &   \\   
      Population (log)    & -34.886$^{*}$     & -35.004$^{*}$ & -34.852$^{*}$ & -35.052$^{*}$\\   
                          & (18.884)          & (19.105)      & (19.209)      & (18.759)\\   
      GDPpc (log)         & -2.813            & -3.202        & -2.888        & -3.150\\   
                          & (8.542)           & (8.738)       & (8.797)       & (8.688)\\   
      R$\&$D ($\%$ GDP)   & -0.172$^{*}$      & -0.181$^{*}$  & -0.180$^{*}$  & -0.174$^{*}$\\   
                          & (0.095)           & (0.095)       & (0.096)       & (0.095)\\   
      Exports ($\%$ GDP)  & 1.991             & 1.891         & 1.970         & 1.906\\   
                          & (1.393)           & (1.407)       & (1.409)       & (1.424)\\   

      \midrule
      \emph{Fixed-effects}\\
      Country             & Yes               & Yes           & Yes           & Yes\\  
      Year                & Yes               & Yes           & Yes           & Yes\\  
      \midrule
      \emph{Fit statistics}\\
      Observations        & 224               & 224           & 224           & 224\\  
      R$^2$               & 0.942             & 0.941         & 0.941         & 0.943\\  
      Within R$^2$        & 0.300             & 0.287         & 0.285         & 0.302\\  
      \midrule \midrule
   \end{tabular}
    \begin{tablenotes}[flushleft]
   \footnotesize
   \item \textit{Notes:} Clustered (Country-Year) standard errors reported in parentheses. 
   All specifications include country and year fixed effects. 
   The dependent variable is the share of labour compensation over Gross Domestic Product, both measured in constant 2015 US Dollars. 
   \textit{Between} captures the between-component of the yearly change in Labour Weighted Fitness, while 
   \textit{Within} captures the within-component of the Labour Weighted Fitness decomposition (both from Equation \ref{eq:dec4}). $\Delta LWF$ denotes the change in the Labour Weighted Fitness index. 
   \textit{GDPpc (log) }is log GDP per capita in US dollars (constant 2015); \textit{Population (log)} is log total population; 
   \textit{R\&D ($\%$ GDP)} is gross domestic expenditure on research and development as a percentage of GDP; 
   \textit{Exports ($\%$ GDP)} is total exports as a percentage of GDP. 
   $^{***}p<0.01$, $^{**}p<0.05$, $^{*}p<0.1$.
   \end{tablenotes}

   \end{threeparttable}
\end{table}

Relatedly, and as a final note on the results presented in this section, it is important to stress that the structural change component is the only one to be significant across all the regression specifications summarised by Equation \ref{eq:model}. As a matter of fact, neither the within-component nor the change in labour-weighted fitness explain any of the outcomes observed, indicating that framing structural change as the movement of labour across industries of different complexity can contribute significantly to explain the current trends in employment growth, wage inequality and functional distribution of income in European countries. 

\subsection{Robustness checks}\label{sec:robustness}

\begin{figure}[!ht]
    \centering
    \caption{Leave-one-out regressions results}
    \includegraphics[width=0.8\linewidth]{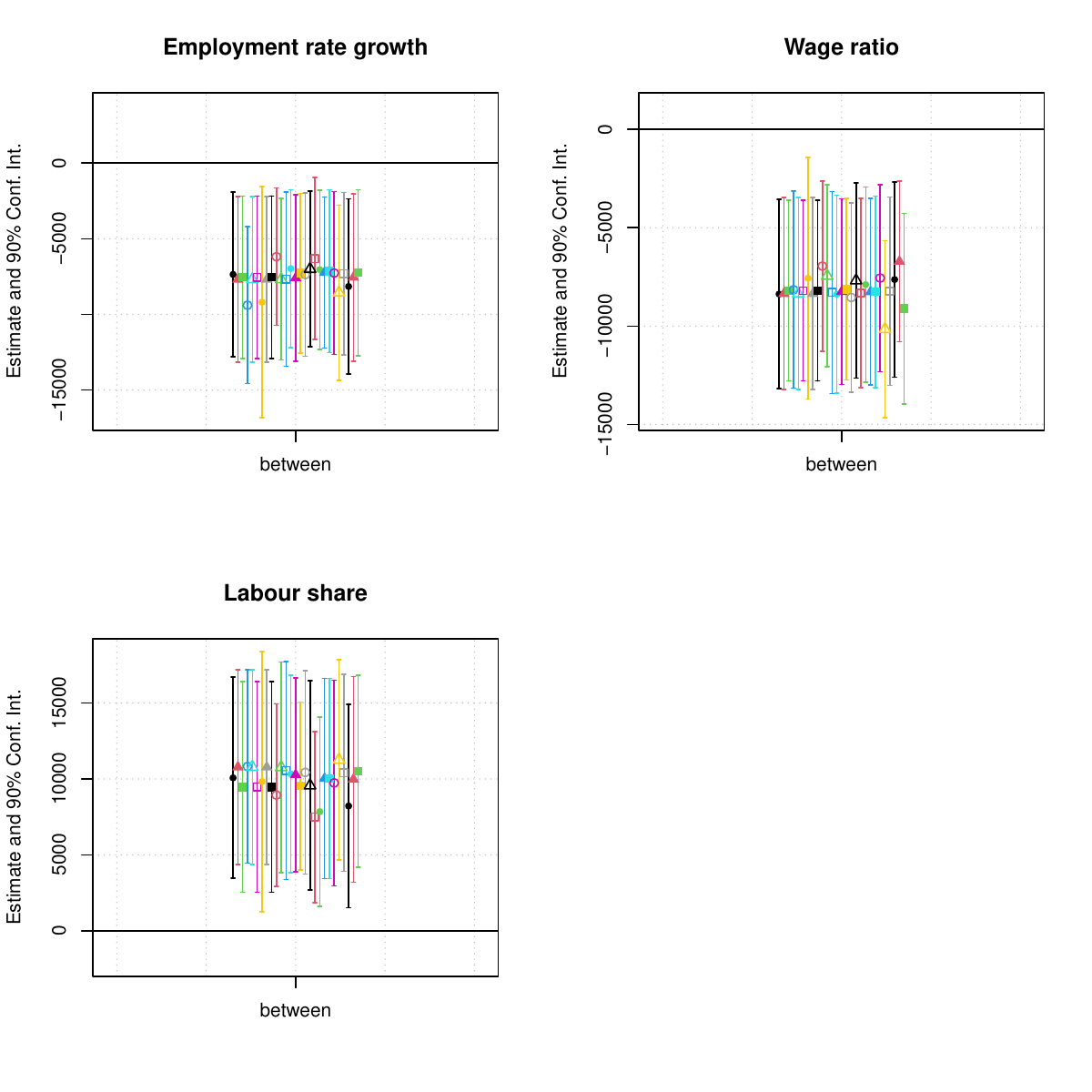}
    \begin{threeparttable}
        \begin{tablenotes}[flushleft]
        \footnotesize
        \item \textit{Notes}: The figure reports the coefficients on the structural change component (the between-term of the Labour-Weighted Fitness decomposition, as defined in Equation \ref{eq:dec4}) obtained from leave-one-out regressions for each of the three outcome variables: yearly employment rate growth (top-left panel), the 9th-to-1st decile wage ratio (top-right panel), and the labour share of the economy (bottom-left panel). Each panel reports the results of N = 30 separate regressions, in each of which one country is excluded from the sample. All specifications follow Equation \ref{eq:model} and include country and year fixed effects, the within-component of the LWF decomposition, log GDP per capita, log population, R\&D expenditure as a share of GDP, and exports as a share of GDP. Standard errors are clustered at the country-year level. Vertical bars represent 90\% confidence intervals around each point estimate. The horizontal black line denotes the zero threshold. Each coloured line corresponds to a regression in which a different country has been excluded from the sample. 
    \end{tablenotes}
\end{threeparttable}
    \label{fig:leaveoneout}
\end{figure}

Given the small number of countries in the sample, our econometric exercise is an inherently small N one, therefore sensitive to changes in individual observations. We test whether the results illustrated in the previous subsection are driven by individual countries exhibiting extreme structural change and/or outcome values. We perform a set of leave-one-out regressions: for each outcome variable (employment growth, wage inequality, and labour share) we run $N = 30$ regressions, excluding in each one a different country. The results of this exercise are provided by Figure \ref{fig:leaveoneout}, which shows the coefficient of the between-component for each of these regressions. As shown in the figure, the coefficient remains always significant, indicating that excluding any country from the sample does not affect the results of any of the models estimated following Equation \ref{eq:model}. 

Additionally, a potential concern is that the signal captured by our structural change component (which is a composite measure combining changes in labour share within industries and their complexity) in explaining the outcomes of interest is driven by either of the elements that feed into the index. In order to test whether it is the case or not, we construct an alternative structural change index, in which the Complexity of industries is replaced by Shannon Entropy. For each time $t$, we compute a Labour Weighted Shannon entropy indicator ($LWS_t$) as follows: 

\begin{equation}
 LWS_t = \sum_{i=1}^n \theta_{i,t}S_{i,t}; \qquad S = H(X) = - \sum_{i = 1}^np(x_i) \, log \, p(x_i).
\label{eq:lws}
\end{equation}

LWS is a measure of sectoral dispersion, as it captures how employment is distributed across industries and therefore the degree of sectoral diversification of the economy. However, it is agnostic about the qualitative content of such diversification: a increase in entropy may reflect a more even distribution of labour across industries, but does not distinguish whether workers are moving into more or less complex industries. In contrast, the Economic Fitness and Complexity framework is explicitly designed to capture the asymmetric distribution of productive capabilities across industries and countries. Accordingly, LWF does not simply reflect whether labour is becoming more dispersed across activities, but rather captures the direction of structural change, identifying whether workers are being reallocated in more complex, capability-intensive sectors.  In other words, while entropy captures diversification and thus the breadth of the productive structure, LWF reflects the direction of structural change by identifying movement towards more complex industries and therefore its position in the hierarchy of productive sophistication.

\begin{table}[!ht]
   \caption{\label{tab:reg_entropy} Labour-weighted entropy regressions}
   \centering
   \footnotesize
   \begin{threeparttable}
   \begin{tabular}{lccc}
      \tabularnewline \midrule \midrule
      Dependent Variables: & Emp. growth ($\%$)  & Ratio $9^{th} / 1^{st}$ dec. & Labour share\\   
       & \multicolumn{3}{c}{FE OLS} \\ 
      Model:               & (1)                              & (2)                          & (3)\\  
      \midrule
      \emph{Variables}\\
      Between (Entropy)           & 5.773                            & 60.682                       & 87.949\\   
                           & (49.927)                         & (37.579)                     & (56.102)\\   
      Within (Entropy)            & -47.473                          & 39.088                       & 94.782\\   
                           & (88.854)                         & (87.878)                     & (127.908)\\   
      Population (log)     & -12.745             & -1.189                         & -29.778\\   
                           & (11.211)                         & (8.642)                       & (20.182)\\   
      GDPpc (log)   & -0.746                           & -3.862                       & -4.433\\   
                           & (4.060)                          & (2.762)                      & (8.187)\\   
      R$\&$D ($\%$ GDP)    & 0.132$^{**}$                 & -0.004                       & -0.175$^{*}$\\   
                           & (0.058)                          & (0.026)                      & (0.097)\\   
      Exports ($\%$ GDP)   & 0.409                & -1.243                       & 1.594\\   
                           & (0.952)                          & (0.972)                      & (1.330)\\   
      \midrule
      \emph{Fixed-effects}\\
      Country          & Yes                              & Yes                          & Yes\\  
      Year                 & Yes                              & Yes                          & Yes\\  
      \midrule
      \emph{Fit statistics}\\
      Observations       & 224                              & 224                          & 224\\  
      R$^2$                           & 0.548                           & 0.905                       & 0.943\\  
      Within R$^2$             & 0.134                           & 0.064                       & 0.304\\  
      \midrule \midrule
   \end{tabular}
   \begin{tablenotes}[flushleft]
   \footnotesize
   \item \textit{Notes:} Clustered (Country-Year) standard errors reported in parentheses. 
   All specifications include country and year fixed effects. The dependent variables are: the growth rate in employment between year $t$ and $t-1$ (1), the wage ratio between the 9$^{th}$ and 1$^{st}$ deciles of the wage distribution (2), and the share of labour compensation over Gross Domestic Product, both measured in constant 2015 US Dollars (3). \textit{Between} captures the between-component of the yearly change in Labour Weighted Fitness, while \textit{Within} captures the within-component of the Labour Weighted Entropy decomposition (both from adapting Equation \ref{eq:dec4}). \textit{GDPpc (log)} is log GDP per capita in US dollars (constant 2015); \textit{Population (log)} is log total population; \textit{R\&D ($\%$ GDP)} is gross domestic expenditure on research and development as a percentage of GDP; \textit{Exports ($\%$ GDP)} is total exports as a percentage of GDP. $^{***}p<0.01$, $^{**}p<0.05$, $^{*}p<0.1$.
   \end{tablenotes}
   \end{threeparttable}
\end{table}

We proceed by regressing the between-component of $LWS$ on employment growth, wage inequality as proxied by the different wage ratios described above, and the labour share. Table \ref{tab:reg_entropy} displays the results of this second robustness exercise, and indicates that the new entropy-based measure of structural change does a very poor job in explaining all three outcomes of interest. This result is informative because it shows how the two measures (fitness and entropy) capture different aspects of structural change. In addition, it corroborates that the explanatory power of the complexity-based structural change measure proposed in this paper does not stem from a generic sectoral compositional change, but from its ability to capture structural upgrading, and is thus is able to capture a strong signal that explains outcomes in employment creation and inequality. In sum, our measure proves to be a solid way to conceptualise structural change as the reallocation of workers across industries with different degrees of complexity.

\section{Conclusions}\label{sec:conclusions}

The empirical analysis conducted in this paper has offered insights on the relationship between structural change -- framed as the reallocation of labour towards more complex industries -- and employment growth, wage inequality, and functional distribution of income. The results indicate that structural change towards complex industries slows down employment creation, due to the lower labour intensity of complex industries. However, the gap between top and bottom wages has shrunk, offering new insights on the inequality reduction brought about by structural change. It must be noted that wage inequality reduces as a result of an increase of the average wages at the first decile of the wage distribution, which in turn is the result of the loss of low-complexity (and likely low pay) jobs by the fittest most diversified countries. The increase in average wage also helps explain the relationship between structural change and the labour share, with the latter increasing when workers move to high-complexity industries. Once more, this effect is not likely to by ascribable to a higher rate of participation to the active workforce, but by the higher salaries paid to workers in countries specialised in complex industries. As a further step to confirm our results, we plan to examine the relationship between the structural change component of the LWF fitness and average wages, which may come in help to rationalise our findings on wage and functional distribution of income. Moreover, in order to confirm the robustness of the results, we are planning to devise an identification strategy based on a shift-share instrument, which appears at the moment the most suitable identification strategy given the nature of our (potentially) endogenous explanatory variable of interest. 

This relationship has been examined introducing several innovative methodological and analytical elements. In the first place, we have introduced a novel method to infer comparative advantages in an unbiased way, in what we believe to be the first empirical application of the BiWCM method proposed by \citet{bruno_inferring_2023}. Secondly, we have constructed measures of Economic Fitness and Complexity that allow the comparison of these two indices over time \citep{mazzilli2024equivalence}, overcoming a known limitation in the use of EC indices in econometric models \citep{caldarola_economic_2024}. Third, we have introduced a new measure -- the Labour-Weighted Fitness -- whose variation allows to be decomposed in such a way as to capture the reallocation of labour across industries with different levels of complexity. This strategy has proved to have significant explanatory power with respect to the creation of employment, the distribution of wages, and the share of labour participation to the aggregate value added. 

Our findings carry several implications for the design of innovation and industrial policy within the European context. First, they suggest that policies aimed exclusively at promoting structural upgrading—through R\&D subsidies, technology adoption incentives, or sector-specific support for knowledge-intensive activities -- may deliver gains in productivity and average wages, but at the cost of slower employment creation. This trade-off cannot be resolved by innovation policy alone: it requires explicit coordination between innovation, industrial, and labour-market policies, in line with recent calls for a more integrated and mission-oriented approach to economic transformation. Second, the heterogeneity in labour absorption across industries that emerges from our analysis indicates that not all forms of structural upgrading are equivalent from a distributive standpoint. The EFC framework, when applied to employment rather than trade data alone, can serve as a diagnostic tool to identify activities that combine high complexity with substantive labour absorption. Third, our finding that wage compression at the bottom of the distribution is largely driven by the disappearance of low-complexity jobs implies that the inclusiveness of structural change is contingent on what happens to displaced workers: in the absence of effective active labour-market policies, reskilling programmes, and social protection, the same process that compresses measured wage inequality may simultaneously generate non-employment and labour-market exit among low-skilled workers, with the observed reduction in inequality reflecting selection out of employment rather than genuine upward mobility. Our findings thus argue for a policy mix that moves beyond stimulating capability accumulation per se, and instead deliberately steers structural transformation towards technological trajectories that conflate sophistication with employment-generating potential, while building the institutional infrastructure -- labour standards, collective bargaining, training systems, and welfare provisions -- needed to ensure that the gains from upgrading are broadly shared.

\newpage
\small
\singlespacing
\clearpage
\phantomsection
\addcontentsline{toc}{chapter}{Bibliography}
\bibliography{labour-fitness}

@Book{McMillan2017a,
  editor    = {McMillan, Margaret and Rodrik, Dani and Sepúlveda, Claudia},
  publisher = {International Food Policy Research Institute (IFPRI)},
  title     = {Structural change, fundamentals, and growth: A framework and case studies},
  year      = {2017},
  address   = {Washington, D.C.},
  doi       = {10.2499/9780896292147},
}

@Article{McMillan2014,
  author   = {Margaret McMillan and Dani Rodrik and Íñigo Verduzco-Gallo},
  journal  = {World Development},
  title    = {Globalization, {Structural} {Change}, and {Productivity} {Growth}, with an {Update} on {Africa}},
  year     = {2014},
  issn     = {0305750X},
  pages    = {11--32},
  volume   = {63},
  doi      = {10.1016/j.worlddev.2013.10.012},
  language = {en},
}

@Article{Vries2015,
  author     = {de Vries, Gaaitzen and Timmer, Marcel and de Vries, Klaas},
  journal    = {The Journal of Development Studies},
  title      = {Structural {Transformation} in {Africa}: {Static} {Gains}, {Dynamic} {Losses}},
  year       = {2015},
  issn       = {0022-0388, 1743-9140},
  number     = {6},
  pages      = {674--688},
  volume     = {51},
  doi        = {10.1080/00220388.2014.997222},
  language   = {en},
  shorttitle = {Structural {Transformation} in {Africa}},
  urldate    = {2019-12-11},
}

@TechReport{Herrendorf2022,
  author      = {Berthold Herrendorf and Richard Rogerson and Ákos Valentinyi},
  institution = {NBER},
  title       = {New evidence on sectoral labour productivity: implications for industrialization and development},
  year        = {2022},
  number      = {29834},
  type        = {Working Paper},
  school      = {National Bureau of Economic Research Working Paper},
}

@InCollection{Herrendorf2014,
  author    = {Herrendorf, Berthold and Rogerson, Richard and Valentinyi, Akos},
  booktitle = {Handbook of Economic Growth},
  publisher = {Elsevier},
  title     = {Growth and structural transformation},
  year      = {2014},
  pages     = {855--941},
  volume    = {2},
}

@Article{Hausmann2009,
  author  = {Hausmann, Ricardo and Hidalgo, César},
  journal = {Proceedings of the National Academy of Sciences of the United States of America},
  title   = {The building blocks of economic complexity},
  year    = {2009},
  number  = {26},
  pages   = {10570--10575},
  volume  = {106},
}

@Article{Cristelli2013,
  author  = {Cristelli, Matthieu and Gabrielli, Andrea and Tacchella, Andrea and Caldarelli, Guido and Pietronero, Luciano},
  journal = {PLoS ONE},
  title   = {Measuring the Intangibles: A Metrics for the Economic Complexity of Countries and Products},
  year    = {2013},
  number  = {8},
  volume  = {8},
}

@Article{Tacchella2012,
  author  = {Andrea Tacchella and Matthieu Cristelli and Guido Caldarelli and Andrea Gabrielli and Luciano Pietronero},
  journal = {Scientific Reports},
  title   = {{A New Metrics for Countries’ Fitness and Products’ Complexity}},
  year    = {2012},
  number  = {723},
  pages   = {1--7},
  volume  = {2},
  doi     = {10.1038/srep00723},
}

@Article{Tacchella2018,
  author   = {Tacchella, Andrea and Mazzilli, Dario and Pietronero, Luciano},
  journal  = {Nature Physics},
  title    = {A dynamical systems approach to gross domestic product forecasting},
  year     = {2018},
  issn     = {1745-2481},
  number   = {8},
  pages    = {861--865},
  volume   = {14},
  doi      = {10.1038/s41567-018-0204-y},
}

@incollection{neffke2026economic,
  author    = {Neffke, Frank and Sbardella, Andrea and Schetter, Ulrich and Tacchella, Andrea},
  title     = {Economic Complexity Analysis},
  booktitle = {The Economy as a Complex Evolving System IV},
  editor    = {Bednar, Jenna and Beinhocker, Eric and del Rio-Chanona, R. Maria and Farmer, J. Doyne and Kaszowska-Mojsa, Joanna and Lafond, Francois and Mealy, Penny and Pangallo, Marco and Pichler, Anton},
  publisher = {Santa Fe Institute Press},
  address   = {Santa Fe, NM},
  year      = {2026}
}

@Article{Castaneda2022,
  author   = {Gonzalo Castañeda and Luciano Pietronero and Juan Romero-Padilla and Andrea Zaccaria},
  journal  = {Structural Change and Economic Dynamics},
  title    = {The complex dynamic of growth: Fitness and the different patterns of economic activity in the medium and long terms},
  year     = {2022},
  issn     = {0954-349X},
  pages    = {231-246},
  volume   = {62},
  doi      = {https://doi.org/10.1016/j.strueco.2022.04.006},
  url      = {https://www.sciencedirect.com/science/article/pii/S0954349X22000601},
}

@Article{Balassa1965,
  author   = {Balassa, Bela},
  journal  = {The Manchester School},
  title    = {{Trade Liberalisation and “Revealed” Comparative Advantage}},
  year     = {1965},
  issn     = {14679957},
  number   = {2},
  pages    = {99--123},
  volume   = {33},
  doi      = {10.1111/j.1467-9957.1965.tb00050.x},
}

@Article{Sbardella2017,
  author   = {Sbardella, Angelica AND Pugliese, Emanuele AND Pietronero, Luciano},
  journal  = {PLOS ONE},
  title    = {Economic development and wage inequality: A complex system analysis},
  year     = {2017},
  number   = {9},
  pages    = {1-26},
  volume   = {12},
  doi      = {10.1371/journal.pone.0182774},
}

@article{napolitano2022green,
  title={Green innovation and income inequality: A complex system analysis},
  author={Napolitano, Lorenzo and Sbardella, Angelica and Consoli, Davide and Barbieri, Nicol{\`o} and Perruchas, Fran{\c{c}}ois},
  journal={Structural Change and Economic Dynamics},
  volume={63},
  pages={224--240},
  year={2022},
  publisher={Elsevier}
}

@article{Hartmann2017,
title = {Linking Economic Complexity, Institutions, and Income Inequality},
journal = {World Development},
volume = {93},
pages = {75-93},
year = {2017},
issn = {0305-750X},
doi = {https://doi.org/10.1016/j.worlddev.2016.12.020},
url = {https://www.sciencedirect.com/science/article/pii/S0305750X15309876},
author = {Dominik Hartmann and Miguel R. Guevara and Cristian Jara-Figueroa and Manuel Aristarán and César A. Hidalgo}
}

@misc{Hartmann2022,
      title={Economic complexity and inequality at the national and regional level}, 
      author={Dominik Hartmann and Flavio L. Pinheiro},
      year={2022},
      eprint={2206.00818},
      archivePrefix={arXiv},
      primaryClass={econ.GN}
}

@Article{Lewis1954,
  author    = {Arthur Lewis},
  journal   = {The Manchester School},
  title     = {{Economic Development with Unlimited Supplies of Labour}},
  year      = {1954},
  issn      = {09654313},
  number    = {8},
  pages     = {139--191},
  volume    = {22},
  address   = {New Haven and London},
  doi       = {10.1080/09654313.2012.680585},
  isbn      = {9781484397480},
  publisher = {Yale University Press},
}

@InCollection{Syrquin1988,
  author    = {Syrquin, Moshe},
  booktitle = {Handbook of Development Economics},
  publisher = {Elsevier},
  title     = {Patterns of structural change},
  year      = {1988},
  chapter   = {07},
  edition   = {1},
  editor    = {Chenery, Hollis and Srinivasan, T.N.},
  pages     = {203-273},
  volume    = {1},
  url       = {https://EconPapers.repec.org/RePEc:eee:devchp:1-07},
}

@Book{Schumpeter1934,
  author    = {Schumpeter, Joseph A.},
  publisher = {Harvard University Press},
  title     = {The Theory of Economic Development: An Inquiry into Profits, Capital, Credit, Interest, and the Business Cycle},
  year      = {1934},
  address   = {Cambridge, MA},
}

@article{Pinheiro2022,
author = {Flavio L. Pinheiro and Pierre-Alexandre Balland and Ron Boschma and Dominik Hartmann},
title = {The dark side of the geography of innovation: relatedness, complexity and regional inequality in Europe},
journal = {Regional Studies},
volume = {Special issue: The dark side of innovation and its geography},
pages = {1-16},
year  = {2022},
publisher = {Routledge},
doi = {10.1080/00343404.2022.2106362},
URL = {https://doi.org/10.1080/00343404.2022.2106362},
eprint = {https://doi.org/10.1080/00343404.2022.2106362}}

@article{Autor2003,
  title={{The skill content of recent technological change: An empirical exploration}},
  author={Autor, David H and Levy, Frank and Murnane, Richard J},
  journal={The Quarterly Journal of Economics},
  volume={118},
  number={4},
  pages={1279--1333},
  year={2003},
  publisher={MIT Press}
}

@techreport{Autor2022,
 title = {The Labor Market Impacts of Technological Change: From Unbridled Enthusiasm to Qualified Optimism to Vast Uncertainty},
 author = {Autor, David},
 institution = {National Bureau of Economic Research},
 type = {Working Paper},
 series = {Working Paper Series},
 number = {30074},
 year = {2022},
 month = {May},
 doi = {10.3386/w30074},
 URL = {http://www.nber.org/papers/w30074},
}

@article{Goos2009,
Author = {Goos, Maarten and Manning, Alan and Salomons, Anna},
Title = {Job Polarization in Europe},
Journal = {American Economic Review},
Volume = {99},
Number = {2},
Year = {2009},
Month = {May},
Pages = {58-63},
DOI = {10.1257/aer.99.2.58},
URL = {https://www.aeaweb.org/articles?id=10.1257/aer.99.2.58}}

@article{Goos2014,
Author = {Goos, Maarten and Manning, Alan and Salomons, Anna},
Title = {Explaining Job Polarization: Routine-Biased Technological Change and Offshoring},
Journal = {American Economic Review},
Volume = {104},
Number = {8},
Year = {2014},
Month = {August},
Pages = {2509-26},
DOI = {10.1257/aer.104.8.2509},
URL = {https://www.aeaweb.org/articles?id=10.1257/aer.104.8.2509}}

@book{Fabricant1942,
  title={Employment in manufacturing, 1899-1939: An analysis of its relation to the volume of production},
  author={Fabricant, Solomon},
  journal={NBER Books},
  year={1942},
  publisher={National Bureau of Economic Research, Inc}
}

@Article{Adam2023,
  author    = {Antonis Adam and Antonios Garas and Marina-Selini Katsaiti and Athanasios Lapatinas},
  journal   = {Economics of Innovation and New Technology},
  title     = {Economic complexity and jobs: an empirical analysis},
  year      = {2023},
  number    = {1},
  pages     = {25-52},
  volume    = {32},
  doi       = {10.1080/10438599.2020.1859751},
  eprint    = { https://doi.org/10.1080/10438599.2020.1859751 },
  publisher = {Routledge},
  url       = {https://doi.org/10.1080/10438599.2020.1859751},
}

@article{HaneWeijman2018,
  title={Returning to work: regional determinants of re-employment after major redundancies},
  author={Emelie Hane-Weijman and Rikard H. Eriksson and Martin Henning},
  journal={Regional Studies},
  year={2018},
  volume={52},
  pages={768 - 780},
  url={https://api.semanticscholar.org/CorpusID:53615559}
}

@article{Dauth2021,
    author = {Dauth, Wolfgang and Findeisen, Sebastian and Suedekum, Jens and Woessner, Nicole},
    title = "{The Adjustment of Labor Markets to Robots}",
    journal = {Journal of the European Economic Association},
    volume = {19},
    number = {6},
    pages = {3104-3153},
    year = {2021},
    month = {05},
    issn = {1542-4766},
    doi = {10.1093/jeea/jvab012},
    url = {https://doi.org/10.1093/jeea/jvab012},
    eprint = {https://academic.oup.com/jeea/article-pdf/19/6/3104/41987038/jvab012.pdf},
}

@Article{Graetz2018,
  author       = {Graetz, Georg and Michaels, Guy},
  year         = {2018},
  journal      = {The Review of Economics and Statistics},
  title        = {{Robots at Work}},
  doi          = {10.1162/rest_a_00754},
  eprint       = {https://direct.mit.edu/rest/article-pdf/100/5/753/1918863/rest\_a\_00754.pdf},
  issn         = {0034-6535},
  number       = {5},
  pages        = {753-768},
  url          = {https://doi.org/10.1162/rest\_a\_00754},
  volume       = {100},
}

@article{mazzilli2024equivalence,
  title={Equivalence between the Fitness-Complexity and the Sinkhorn-Knopp algorithms},
  author={Mazzilli, Dario and Mariani, Manuel Sebastian and Morone, Flaviano and Patelli, Aurelio},
  journal={Journal of Physics: Complexity},
  volume = {5},
  issue = {1},
  pages = {015010},
  year={2024}
}

@Article{krantz2018,
AUTHOR = {Krantz, Ruben and Gemmetto, Valerio and Garlaschelli, Diego},
TITLE = {Maximum-Entropy Tools for Economic Fitness and Complexity},
JOURNAL = {Entropy},
VOLUME = {20},
YEAR = {2018},
NUMBER = {10},
ARTICLE-NUMBER = {743},
URL = {https://www.mdpi.com/1099-4300/20/10/743},
ISSN = {1099-4300},
DOI = {10.3390/e20100743}
}

@article{Vollrath1991,
 ISSN = {00432636},
 URL = {http://www.jstor.org/stable/40439943},
 author = {Thomas L. Vollrath},
 journal = {Weltwirtschaftliches Archiv},
 number = {2},
 pages = {265--280},
 publisher = {Springer},
 title = {A Theoretical Evaluation of Alternative Trade Intensity Measures of Revealed Comparative Advantage},
 urldate = {2023-02-08},
 volume = {127},
 year = {1991}
}

@article{bruno_inferring_2023,
doi = {10.1088/2632-072X/ad1411},
url = {https://doi.org/10.1088/2632-072X/ad1411},
year = {2023},
month = {dec},
publisher = {IOP Publishing},
volume = {4},
number = {4},
pages = {045011},
author = {Bruno, Matteo and Mazzilli, Dario and Patelli, Aurelio and Squartini, Tiziano and Saracco, Fabio},
title = {Inferring comparative advantage via entropy maximization},
journal = {Journal of Physics: Complexity}
}

@article{Kuznets1955,
 ISSN = {00028282},
 URL = {http://www.jstor.org/stable/1811581},
 author = {Simon Kuznets},
 journal = {The American Economic Review},
 number = {1},
 pages = {1--28},
 publisher = {American Economic Association},
 title = {Economic Growth and Income Inequality},
 urldate = {2023-10-01},
 volume = {45},
 year = {1955}
}

@article{Antonelli2020,
title = {Income inequality in the knowledge economy},
journal = {Structural Change and Economic Dynamics},
volume = {55},
pages = {153-164},
year = {2020},
issn = {0954-349X},
doi = {https://doi.org/10.1016/j.strueco.2020.07.003},
url = {https://www.sciencedirect.com/science/article/pii/S0954349X2030374X},
author = {Cristiano Antonelli and Matteo Tubiana}
}

@InCollection{Freire2021,
  author    = {Clovis Freire},
  booktitle = {{New Perspectives on Structural Change}},
  publisher = {Oxford University Press},
  title     = {{Economic Complexity Perspectives on Structural Change}},
  year      = {2021},
  pages     = {188--214},
  doi       = {10.1093/oso/9780198850113.003.0010},
}

@article{Acemoglu2022,
	title = {Tasks, {Automation}, and the {Rise} in {U}.{S}. {Wage} {Inequality}},
	volume = {90},
	issn = {0012-9682},
	url = {https://www.econometricsociety.org/doi/10.3982/ECTA19815},
	doi = {10.3982/ECTA19815},
	language = {en},
	number = {5},
	urldate = {2023-09-12},
	journal = {Econometrica},
	author = {Acemoglu, Daron and Restrepo, Pascual},
	year = {2022},
	pages = {1973--2016},
}

@book{Brynjolfsson2011,
  title={Race against the machine: How the digital revolution is accelerating innovation, driving productivity, and irreversibly transforming employment and the economy},
  author={Brynjolfsson, Erik and McAfee, Andrew},
  year={2011},
  publisher={Brynjolfsson and McAfee}
}

@article{Frey2017,
  title={{The future of employment: How susceptible are jobs to computerisation?}},
  author={Frey, Carl Benedikt and Osborne, Michael A},
  journal={Technological Forecasting and Social Change},
  volume={114},
  pages={254--280},
  year={2017},
  publisher={Elsevier}
}

@Article{Cimoli1995,
  author   = {Cimoli, Mario and Dosi, Giovanni},
  journal  = {Journal of Evolutionary Economics},
  title    = {Technological paradigms, patterns of learning and development: An introductory roadmap},
  year     = {1995},
  issn     = {1432-1386},
  number   = {3},
  pages    = {243--268},
  volume   = {5},
  doi      = {10.1007/BF01198306},
  refid    = {Cimoli1995},
  url      = {https://doi.org/10.1007/BF01198306},
}

@Book{Hirschman1958,
  author    = {Albert O Hirschman},
  publisher = {Yale University Press},
  title     = {The {Strategy} of {Economic} {Development}},
  year      = {1958},
  address   = {New Haven},

}

@article{Teece1994,
  title={Understanding corporate coherence: {T}heory and evidence},
  author={Teece, David J and Rumelt, Richard and Dosi, Giovanni and Winter, Sidney},
  journal={Journal of Economic Behavior \& Organization},
  volume={23},
  number={1},
  pages={1--30},
  year={1994},
  publisher={Elsevier}
}

@TechReport{Hausmann2006,
  author      = {Ricardo Hausmann and Benjamin Klinger},
  institution = {Centre for International Development, Harvard University},
  title       = {Structural Transformation and Patterns of Comparative Advantage in the Product Space},
  year        = {2006},
  month       = sep,
  number      = {RWP06-041},
  type        = {CID Working Paper},
  doi         = {http://dx.doi.org/10.2139/ssrn.939646},
}

@article{Bustos2012,
	title = {The {Dynamics} of {Nestedness} {Predicts} the {Evolution} of {Industrial} {Ecosystems}},
	volume = {7},
	issn = {19326203},
	doi = {10.1371/journal.pone.0049393},
	number = {11},
	journal = {PLoS ONE},
	author = {Bustos, Sebastián and Gomez, Charles and Hausmann, Ricardo and Hidalgo, César A.},
	year = {2012},
	pmid = {23185326},
	pages = {1--8},
}

@Article{Patelli2023,
  author   = {Patelli, Aurelio and Napolitano, Lorenzo and Cimini, Giulio and Pugliese, Emanuele and Gabrielli, Andrea},
  journal  = {Scientific Reports},
  title    = {Capability accumulation patterns across economic, innovation, and knowledge-production activities},
  year     = {2023},
  issn     = {2045-2322},
  number   = {1},
  pages    = {12988},
  volume   = {13},
  doi      = {10.1038/s41598-023-29979-x},
  refid    = {Patelli2023},
  url      = {https://doi.org/10.1038/s41598-023-29979-x},
}

@article{caldarola_economic_2024,
	title = {Economic complexity and the sustainability transition: a review of data, methods, and literature},
	volume = {5},
	copyright = {CC0 1.0 Universal Public Domain Dedication},
	issn = {2632-072X},
	shorttitle = {Economic complexity and the sustainability transition},
	url = {https://iopscience.iop.org/article/10.1088/2632-072X/ad4f3d},
	doi = {10.1088/2632-072X/ad4f3d},
	language = {en},
	number = {2},
	urldate = {2024-06-17},
	journal = {Journal of Physics: Complexity},
	author = {Caldarola, Bernardo and Mazzilli, Dario and Napolitano, Lorenzo and Patelli, Aurelio and Sbardella, Angelica},
	month = jun,
	year = {2024},
	pages = {022001},
}

@article{Loturco2022,
title = {The knowledge and skill content of production complexity},
journal = {Research Policy},
volume = {51},
number = {8},
pages = {104059},
year = {2022},
note = {Special Issue on Economic Complexity},
issn = {0048-7333},
doi = {https://doi.org/10.1016/j.respol.2020.104059},
url = {https://www.sciencedirect.com/science/article/pii/S0048733320301372},
author = {Alessia {Lo Turco} and Daniela Maggioni}
}

@book{Picketty2014,
  title={Capital in the twenty-first century},
  author={Piketty, Thomas},
  year={2014},
  publisher={Harvard University Press}
}

@article{Picketty2018,
  title={Distributional national accounts: methods and estimates for the United States},
  author={Piketty, Thomas and Saez, Emmanuel and Zucman, Gabriel},
  journal={The Quarterly Journal of Economics},
  volume={133},
  number={2},
  pages={553--609},
  year={2018},
  publisher={Oxford University Press}
}

@article{Haltiwanger2024,
Author = {Haltiwanger, John and Hyatt, Henry R. and Spletzer, James R.},
Title = {Rising Top, Falling Bottom: Industries and Rising Wage Inequality},
Journal = {American Economic Review},
Volume = {114},
Number = {10},
Year = {2024},
Month = {October},
Pages = {3250–83},
DOI = {10.1257/aer.20221574},
URL = {https://www.aeaweb.org/articles?id=10.1257/aer.20221574}}

@article{Karabarbounis2024,
Author = {Karabarbounis, Loukas},
Title = {Perspectives on the Labor Share},
Journal = {Journal of Economic Perspectives},
Volume = {38},
Number = {2},
Year = {2024},
Month = {May},
Pages = {107–36},
DOI = {10.1257/jep.38.2.107},
URL = {https://www.aeaweb.org/articles?id=10.1257/jep.38.2.107}}

@article{Mcnerney2025,
  title={Bridging the short-term and long-term dynamics of economic structural change},
  author={McNerney, James and Li, Yang and Gomez-Lievano, Andres and Neffke, Frank},
  journal={Nature Communications},
  volume={16},
  number={1},
  pages={10225},
  year={2025},
  publisher={Nature Publishing Group UK London}
}

@article{autor_polarization_2006,
	title = {The {Polarization} of the {U}.{S}. {Labor} {Market}},
    journal = {{A}merican {E}conomic {R}eview},
	volume = {96},
    issue = {2},
    pages = {189-194},
	language = {en},
	number = {2},
	author = {Autor, David H and Katz, Lawrence F and Kearney, Melissa S},
	year = {2006},
}

@article{acemoglu_automation_2026,
	title = {Automation and {Polarization}},
	issn = {0022-3808, 1537-534X},
	url = {https://www.journals.uchicago.edu/doi/10.1086/739330},
	doi = {10.1086/739330},
	language = {en},
	urldate = {2026-02-05},
	journal = {Journal of Political Economy},
	author = {Acemoglu, Daron and Loebbing, Jonas},
	month = jan,
	year = {2026},
	pages = {000--000},
}

@article{teixeira_scientific_2026,
	title = {Scientific and industrial specialisation, structural change and economic growth: {Global} evidence},
	volume = {55},
	issn = {00487333},
	shorttitle = {Scientific and industrial specialisation, structural change and economic growth},
	url = {https://linkinghub.elsevier.com/retrieve/pii/S0048733325002124},
	doi = {10.1016/j.respol.2025.105383},
	language = {en},
	number = {2},
	urldate = {2026-01-15},
	journal = {Research Policy},
	author = {Teixeira, Aurora A.C. and Pinto, Alexandre},
	month = mar,
	year = {2026},
	pages = {105383},
}

@article{inoua2023,
title = {A simple measure of economic complexity},
journal = {Research Policy},
volume = {52},
number = {7},
pages = {104793},
year = {2023},
issn = {0048-7333},
doi = {10.1016/j.respol.2023.104793},
author = {Sabiou Inoua}
}

@article{vandam2022,
title = {Variety, complexity and economic development},
journal = {Research Policy},
volume = {51},
number = {8},
pages = {103949},
year = {2022},
note = {Special Issue on Economic Complexity},
issn = {0048-7333},
doi = {https://doi.org/10.1016/j.respol.2020.103949},
url = {https://www.sciencedirect.com/science/article/pii/S0048733320300299},
author = {Alje van Dam and Koen Frenken}
}

@incollection{matsuyama2008,
	address = {London},
	title = {Structural {Change}},
	isbn = {978-1-349-95121-5},
	url = {https://link.springer.com/10.1057/978-1-349-95121-5\_1775-2},
	doi = {10.1057/978-1-349-95121-5\_1775-2},
	language = {en},
	urldate = {2023-09-12},
	booktitle = {The {New} {Palgrave} {Dictionary} of {Economics}},
	publisher = {Palgrave Macmillan UK},
	author = {Matsuyama, Kiminori},
	editor = {{Palgrave Macmillan}},
	year = {2008},
	pages = {1--6},
}

@incollection{hartmann2024economic,
  title={Economic complexity and inequality at the national and regional levels},
  author={Hartmann, Dominik and Pinheiro, Fl{\'a}vio L},
  booktitle={Routledge International Handbook of Complexity Economics},
  pages={551--566},
  year={2024},
  publisher={Routledge}
}

@article{verspagen1991new,
  title={A new empirical approach to catching up or falling behind},
  author={Verspagen, Bart},
  journal={Structural change and economic dynamics},
  volume={2},
  number={2},
  pages={359--380},
  year={1991},
  publisher={North-Holland}
}

@book{cimoli_industrial_2009,
	title = {Industrial {Policy} and {Development}: {The} {Political} {Economy} of {Capabilities} {Accumulation}},
	url = {https://EconPapers.repec.org/RePEc:oxp:obooks:9780199235278},
	publisher = {Oxford University Press},
	editor = {Cimoli, Mario and Dosi, Giovanni and Stiglitz, Joseph},
	year = {2009},
}

@incollection{Forster2015,
title = {Chapter 19 - Cross-Country Evidence of the Multiple Causes of Inequality Changes in the OECD Area},
editor = {Anthony B. Atkinson and François Bourguignon},
series = {Handbook of Income Distribution},
publisher = {Elsevier},
volume = {2},
pages = {1729-1843},
year = {2015},
booktitle = {Handbook of Income Distribution},
issn = {1574-0056},
doi = {https://doi.org/10.1016/B978-0-444-59429-7.00020-0},
url = {https://www.sciencedirect.com/science/article/pii/B9780444594297000200},
author = {Michael F. Förster and István György Tóth}
}

@techreport{cosentino_structural_2023,
	institution = {United Nations University (UNU), Maastricht Economic and Social Research Institute on Innovation and Technology (UNU-MERIT), Maastricht},
	series = {{UNU}-{MERIT} {Working} {Papers}},
	title = {Structural change and income inequality: a meta-analysis},
	language = {en},
	number = {2023-46},
	publisher = {UNU-MERIT Working Papers},
	author = {Cosentino De la Vega, Rafael},
	year = {2023},
}

@incollection{kanbur_income_2000,
	title = {Income distribution and development},
	volume = {1},
	language = {en},
	booktitle = {Handbook of {Income} {Distribution}},
	publisher = {North Holland},
	author = {Kanbur, Ravi},
	collaborator = {Atkinson, Anthony Barnes and Bourguignon, François},
	year = {2000},
}

@book{diodato_handbook_2024,
	publisher = {Publications Office of the European Union, Luxembourg},
	title = {Handbook of {Economic} {Complexity} for {Policy}},
	volume = {JRC138666},
	url = {https://data.europa.eu/doi/10.2760/9006857},
	doi = {10.2760/9006857},
	author = {Diodato, Dario and Napolitano, Lorenzo and Pugliese, Emanuele and Tacchella, Andrea},
	year = {2024}
}

@article{baumol1967macroeconomics,
  title={Macroeconomics of unbalanced growth: the anatomy of urban crisis},
  author={Baumol, William J},
  journal={The American economic review},
  volume={57},
  number={3},
  pages={415--426},
  year={1967},
  publisher={JSTOR}
}

@article{atkinson2006panel,
  title={The panel-of-countries approach to explaining income inequality: an interdisciplinary research agenda},
  author={Atkinson, Anthony B and Brandolini, Andrea},
  journal={Mobility and inequality: Frontiers of research in sociology and economics},
  pages={400--448},
  year={2006},
  publisher={Stanford, Stanford University Press}
}

@incollection{hellier2013inequality,
  title={Inequality, growth and welfare: The main links},
  author={Hellier, Jo{\"e}l and Lambrecht, St{\'e}phane},
  booktitle={Growing Income Inequalities: Economic Analyses},
  pages={274--311},
  year={2013},
  publisher={Springer}
}

@techreport{caldarola2022structural,
  title={Structural change (s) in Ghana: A comparison between the trade, formal and informal sectors},
  author={Caldarola, Bernardo},
  series={Laboratory of Economics and Management (LEM) Working Paper Series},
  institution={Sant'Anna School of Advanced Studies},
  number={2022/36},
  year={2022}
}

@article{domanski2003industrial,
  title={{Industrial change and foreign direct investment in the postsocialist economy: the case of Poland}},
  author={Doma{\'n}ski, Boles{\l}aw},
  journal={European Urban and Regional Studies},
  volume={10},
  number={2},
  pages={99--118},
  year={2003},
  publisher={Sage Publications}
}

@techreport{OECD2025PolandFDI,
  author      = {{OECD}},
  title       = {Strengthening FDI and SME Linkages in Poland},
  institution = {OECD},
  year        = {2025},
  url         = {https://www.oecd.org/en/publications/strengthening-fdi-and-sme-linkages-in-poland_cb08eb93-en.html}
}

@incollection{abramovitz1995elements,
  title={The elements of social capability},
  author={Abramovitz, Moses},
  booktitle={Social capability and long-term economic growth},
  pages={19--47},
  year={1995},
  publisher={Springer}
}

@article{hausmann2003economic,
  title={Economic development as self-discovery},
  author={Hausmann, Ricardo and Rodrik, Dani},
  journal={Journal of development Economics},
  volume={72},
  number={2},
  pages={603--633},
  year={2003},
  publisher={Elsevier}
}

@article{aufiero2024mapping,
  title={Mapping job fitness and skill coherence into wages: an economic complexity analysis},
  author={Aufiero, Sabrina and De Marzo, Giordano and Sbardella, Angelica and Zaccaria, Andrea},
  journal={Scientific Reports},
  volume={14},
  number={1},
  pages={11752},
  year={2024},
  publisher={Nature Publishing Group UK London}
}

@article{galbraith2025kuznets,
  title={Kuznets at 70 The enduring significance of a curve and a hypothesis},
  author={Galbraith, James and Kanbur, Ravi and Sen, Kunal and Sumner, Andy},
  journal={Structural Change and Economic Dynamics},
  year={2025},
  volume={77},
  pages={248-257},
  publisher={Elsevier}
}

@article{galbraith2011inequality,
  author  = {Galbraith, James K.},
  title   = {Inequality and Economic and Political Change: A Comparative Perspective},
  journal = {Cambridge Journal of Regions, Economy and Society},
  volume  = {4},
  number  = {1},
  pages   = {13--27},
  year    = {2011},
  doi     = {10.1093/cjres/rsq014}
}

@article{stockhammer2012have,
  title={Why have wage shares fallen? A panel analysis of the determinants of functional income distribution},
  author={Stockhammer, Engelbert},
  journal={Conditions of Work and Employment Series, Geneva: ILO},
  volume={15},
  year={2012}
}

@article{mishel2021identifying,
  title={Identifying the policy levers generating wage suppression and wage inequality},
  author={Mishel, Lawrence and Bivens, Josh},
  journal={Economic Policy Institute},
  volume={13},
  number={1},
  pages={5--12},
  year={2021}
}

@article{feenstra1996globalization,
  title={Globalization, Outsourcing, and Wage Inequality},
  author={Feenstra, Robert C and Hanson, Gordon H},
  journal={The American Economic Review},
  volume={86},
  number={2},
  pages={240--245},
  year={1996},
  publisher={JSTOR}
}

@article{timmer2014slicing,
  title={Slicing up global value chains},
  author={Timmer, Marcel P and Erumban, Abdul Azeez and Los, Bart and Stehrer, Robert and De Vries, Gaaitzen J},
  journal={Journal of Economic Perspectives},
  volume={28},
  number={2},
  pages={99--118},
  year={2014},
  publisher={American Economic Association 2014 Broadway, Suite 305, Nashville, TN 37203-2418}
}

@article{rodrik2013structural,
  title={Structural change, fundamentals, and growth: an overview},
  author={Rodrik, Dani and others},
  journal={Institute for Advanced Study},
  volume={23},
  pages={1--12},
  year={2013}
}

@article{rodrik2016premature,
  author  = {Rodrik, Dani},
  title   = {Premature Deindustrialization},
  journal = {Journal of Economic Growth},
  volume  = {21},
  number  = {1},
  pages   = {1--33},
  year    = {2016},
  doi     = {10.1007/s10887-015-9122-3}
}

@article{jorgenson2011structural,
  title={Structural change in advanced nations: a new set of stylised facts},
  author={Jorgenson, Dale W and Timmer, Marcel P},
  journal={Scandinavian Journal of Economics},
  volume={113},
  number={1},
  pages={1--29},
  year={2011},
  publisher={Wiley Online Library}
}

@article{cirillo2018job,
  title={Job polarization in European industries},
  author={Cirillo, Valeria},
  journal={International Labour Review},
  volume={157},
  number={1},
  pages={39--63},
  year={2018},
  publisher={Wiley Online Library}
}

@article{autor2013china,
  title={The China syndrome: Local labor market effects of import competition in the United States},
  author={Autor, David H and Dorn, David and Hanson, Gordon H},
  journal={American economic review},
  volume={103},
  number={6},
  pages={2121--2168},
  year={2013},
  publisher={American Economic Association}
}

@article{stockhammer2017determinants,
  author  = {Stockhammer, Engelbert},
  title   = {Determinants of the Wage Share: A Panel Analysis of Advanced and Developing Economies},
  journal = {British Journal of Industrial Relations},
  volume  = {55},
  number  = {1},
  pages   = {3--33},
  year    = {2017},
  doi     = {10.1111/bjir.12165}
}

@techreport{kucera2012structure,
  author      = {Kucera, David and Roncolato, Leanne},
  title       = {Structure Matters: Sectoral Drivers of Growth and the Labour Productivity-Employment Relationship},
  institution = {International Labour Organization},
  type        = {ILO Working Papers},
  number      = {994717343402676},
  year        = {2012}
}

@techreport{stansbury2020declining,
  title={The declining worker power hypothesis: An explanation for the recent evolution of the American economy},
  author={Stansbury, Anna and Summers, Lawrence H},
  year={2020},
  institution={National Bureau of Economic Research}
}

@article{karabarbounis2014global,
  title={The global decline of the labor share},
  author={Karabarbounis, Loukas and Neiman, Brent},
  journal={The Quarterly journal of economics},
  volume={129},
  number={1},
  pages={61--103},
  year={2014},
  publisher={Oxford University Press}
}

@article{hubmer2026,
	title = {Not a {Typical} {Firm}: {Capital}–{Labor} {Substitution} and {Firms}' {Labor} {Shares}},
	volume = {18},
	issn = {1945-7707, 1945-7715},
	shorttitle = {Not a {Typical} {Firm}},
	url = {https://pubs.aeaweb.org/doi/10.1257/mac.20230325},
	doi = {10.1257/mac.20230325},
	language = {en},
	number = {2},
	urldate = {2026-04-13},
	journal = {American Economic Journal: Macroeconomics},
	author = {Hubmer, Joachim and Restrepo, Pascual},
	month = apr,
	year = {2026},
	pages = {34--71},
}

@article{paredes2021automation,
  title={Automation and robotics in mining: Jobs, income and inequality implications},
  author={Paredes, Dusan and Fleming-Mu{\~n}oz, David},
  journal={The Extractive Industries and Society},
  volume={8},
  number={1},
  pages={189--193},
  year={2021},
  publisher={Elsevier}
}

@article{Dimova2019,
  title={{The structural determinants of the labor share in Europe}},
  author={Dimova, Dilyana},
  year={2019},
  journal={IMF Working Papers},
  volume={2019},
  number={067},
  
}

@article{Arpaia2009,
  title={{Understanding labour income share dynamics in Europe}},
  author={Arpaia, Alfonso and P{\'e}rez, Esther and Pichelmann, Karl},
  year={2009},
  volume={379},
  journal={Working paper, European Commission, European Economy Paper},
}

@article{Mishra2020,
  title={Economic complexity and the globalization of services},
  author={Mishra, Saurabh and Tewari, Ishani and Toosi, Siavash},
  journal={Structural Change and Economic Dynamics},
  volume={53},
  pages={267--280},
  year={2020},
  publisher={Elsevier}
}
\newpage

\appendix
\normalsize
\onehalfspacing

\renewcommand{\thetable}{A\arabic{table}}
\setcounter{table}{0}

\renewcommand{\thefigure}{A\arabic{figure}}
\setcounter{figure}{0}

\section{Data reconstruction}\label{sec:data-rec}

The Eurostat's Structural Business Survey data used to compute the employment based EFC measures used in this paper are one of the most comprehensive and detailed data sources of employment in European industries, classified at a fine level of disaggregation (NACE 4 digit). However, the dataset presents a large number of missing values, which prevent the construction of a specialisation matrix as explained in Section \ref{sec:biwcm}. In fact, the BiWCM and EFC algorithm require both a matrix with complete information. 

In order to address this shortcoming in the Structural Business Survey data described in Section \ref{sec:sbs}, we have pursued, compared, and validated seven different interpolation strategies, in order to retrieve a good approximation of the missing data. We have attempted the reconstruction of the SBS data following seven alternative strategies, applied to the country-industry time series having a maximum seven missing data points.\footnote{In the case of fully missing series, the country-industry couple has been excluded. In case of eight out of nine missing data points, the value has been assumed to be constant across the time series, extrapolating at constant values from the only available data point in the series.} Below we summarise the seven distinct strategies:

\begin{enumerate}
    \item Internal missing values (those after the first available data point, and before the last) are interpolated using linear interpolation, while external missing values (before and after, respectively, the first and last data point available) have been extrapolated backwards taking the first available value as constant;
    \item Internal missing values are interpolated using linear interpolation, while external missing values are extrapolated using the closest growth rate available, obtained by after interpolating internal missing values;
    \item Internal missing values are interpolated using linear interpolation, while external missing values are extrapolated using average growth rate applied to the fist and last data points available after interpolating. This strategy produced negative values, which were constrained to 0;
    \item Internal missing values are interpolated using linear interpolation, while external missing values are extrapolated using the first available growth rate by filling up/downwards available growth rates;
    \item Internal missing values are interpolated using linear interpolation, while external missing values are extrapolated using moving average of growth rates, constructed using a rolling window of 3 years. The first two years are always `NA`; second year is filled with moving average for t=2, and first year is just the growth with respect to the previous year;
    \item Interpolate and extrapolate with linear fit computed using the available data points in each country-industry series;
    \item Internal missing values are interpolated using linear interpolation, while external missing values are extrapolated using linear fit.

\end{enumerate}

Each strategy is compared and validated using information from the complete country-industry time series. After filtering only the complete series from the dataset, we have created five different validation subsamples by randomising missing values in the complete dataset, imposing the following conditions:
\begin{itemize}
    \item Reflect NA frequency in original data (14\% of data missing)
    \item Reflect frequency in original data $\pm$ 5\%,
    \item Reflect frequency in original data $\pm$ 10\%, 
    \item Impose 50\% of missing values.

\end{itemize}

\begin{table}[!htbp]
\caption{Mean Absolute Errors}
\small
\centering

\begin{threeparttable}

\begin{tabular}{lccccccc}
\tabularnewline \toprule
               & \multicolumn{7}{c}{Reconstruction strategy} \\ 
\cmidrule{2-8}
Subsample      & 1 & 2 & 3 & 4 & 5 & 6 & 7 \\ 
\midrule
NA 9\%   & \textbf{174.96} & 189.05 & 179.47 & 192.23 & 189.44 & 222.55 & 197.73 \\
NA 14\%  & \textbf{214.95} & 227.60 & 215.10 & 230.20 & 227.45 & 273.22 & 240.86 \\
NA 19\%  & \textbf{254.61} & 294.14 & 267.06 & 297.80 & 292.47 & 323.64 & 286.27 \\
NA 24\%  & \textbf{311.79} & 329.90 & 315.13 & 331.72 & 325.46 & 395.41 & 355.26 \\
NA 50\%  & \textbf{692.86} & 788.71 & 749.16 & 796.20 & 784.38 & 819.18 & 777.39 \\ 
\bottomrule
\end{tabular}

\begin{tablenotes}[flushleft]
\footnotesize
\item Notes: The table reports mean absolute errors (MAE) across different levels of missing data (NA). 
Bold values indicate the lowest MAE within each subsample.
\end{tablenotes}

\end{threeparttable}

\label{tab:mae_interpolation}
\end{table}

In order to choose the data reconstruction strategy that best approximates the observed data, we compare the Mean Absolute Error of each strategy across the 5 validation samples. As shown by Table \ref{tab:mae_interpolation}, the strategy that minimises the prediction error across the different validation subsamples is strategy number 1. Therefore, the final employment dataset used to construct the specialisation matrix interpolates internal missing values linearly, and extrapolates external values taking the first and last available data point as constant, both backward and forward. According to our test, this strategy is the one that is most likely replicate the real data.

\section{Methodological notes}

\subsection{Differences between Revealed and Inferred Comparative Advantage}\label{sec:ICA}
The primary rationale for utilising a Comparative Advantage index is to establish a measure that isn't contingent on the relative strength of diverse countries and industries, which are inherently incomparable in terms of size. Revealed Comparative Advantage (RCA) and Inferred Comparative Advantage (ICA) share the same conceptual core: both assess whether the observed extensive quantity, a \emph{volume}, $V_{ci}$ of actor $c$ in activity $i$ exceeds an \emph{expected} value under a benchmark. In this work, the volume is the labour force, the actor is a country, and the activity is an industry (NACE 4 digits). In general terms, a comparative advantage index can be cast as an observed-to-expected ratio:

\begin{equation}
    \mathrm{CA}_{ci}=\frac{V_{ci}}{\mathbb{E}[V_{ci}]}\,
\end{equation}

Where observed term is the share of country $c$ in industry $i$, and the expected term is -- in general terms -- an expectation of volume in industry $i$ based on the whole sample of countries. The difference between RCA rests in the way in which the expectation $\mathbb{E}[V_{cp}]$ is constructed -- namely on the constraints imposed on the expected value of volume.

The traditional RCA measure proposed by \citet{Balassa1965}, specifically, contrasts the actual volume of a country with an anticipated value derived under the assumption of equalising country and sector sizes~\citep{Vollrath1991,krantz2018}. In other words, the share of country $c$ in industry $i$ is compared to the average share of all countries in the same industry $i$: 
\begin{equation}
    \mathrm{RCA}_{ci} =\frac{V_{ci}/\sum_{i'}V_{ci'}}{\left(\sum_{c'}V_{c'i}\right)/\sum_{c'i'}V_{c'i'}}\,,
\end{equation}
which is equivalent to comparing $V_{ci}$ to a factorised expectation based on marginal totals: let $V_c=\sum_{i}V_{cp}$, $V_p=\sum_{c}V_{ci}$, and $V=\sum_{c,i}V_{ci}$. Then
\begin{equation}
    \mathrm{RCA}_{ci}=\frac{V_{ci}}{\mathbb{E}_{\mathrm{RCA}}[V_{ci}]}
\qquad\text{with}\qquad
\mathbb{E}_{\mathrm{RCA}}[V_{ci}]=\frac{V_c\,V_i}{V}\,.
\end{equation}
Therefore, RCA implicitly adopts an \emph{independence} baseline where the expected intensity in $(c,i)$ is determined solely by the overall scale of $c$ and the overall prevalence of $i$.
This assumption is generally considered reasonable when the volumes are of similar magnitude and no correlation is present in the system.

ICA \citep{bruno_inferring_2023} avoids this naive assumption, which may not apply to all types of data. ICA retains the same observed-to-expected structure but replaces the implicit baseline with an explicit probabilistic null model derived by maximum entropy. 
Specifically, the expectation $\mathbb{E}[V_{ci}]$ is computed under the \emph{Bipartite Weighted Configuration Model} (BiWCM), \textit{i.e.} the maximum-entropy ensemble of weighted bipartite networks that is maximally random subject to reproducing the total marginal volumes as constraints (\textit{i.e.} the node strengths, a.k.a. the row and column sums) in expectation. 
Denoting the BiWCM-implied expected weight by $\mathbb{E}_{\mathrm{BiWCM}}[V_{ci}]$, ICA is defined as
\begin{equation}
\mathrm{ICA}_{ci}=\frac{V_{ci}}{\mathbb{E}_{\mathrm{BiWCM}}[V_{ci}]}\,.
\end{equation}

The key methodological analogy is that both RCA and ICA quantify comparative advantage as a deviation from an expected value anchored to aggregate information. The key methodological difference is that RCA uses a closed-form expectation $\mathbb{E}_{\mathrm{RCA}}[V_{ci}]=V_cV_i/V$ implied by the volume independence assumption, whereas ICA uses the maximum-entropy expectation $\mathbb{E}_{\mathrm{BiWCM}}[V_{ci}]$, obtained from an explicit entropy-maximising (equivalently, maximum-likelihood) null model consistent with the imposed constraints. This approach naturally supports uncertainty quantification and statistical validation when needed, while minimising statistical bias.

Finally, the comparative advantage index is binarised whether it is larger or smaller of a threshold.
The value is naturally set to 1, stating that we assume a comparative advantage whenever a country in a product does more than expected by the null model.
Therefore we have:
\begin{equation}
    M_{c,i} = \begin{cases}
    1 & \mathrm{ICA}_{c,i}\geq 1\\
    0 & \text{otherwise}
    \end{cases}
\end{equation}

Economically, ICA can be interpreted as asking whether a country employs \emph{more labour in a given industry than would be expected purely from its overall size and the global importance of that industry}, once the structural constraints of the entire country–industry system are taken into account. In this sense, ICA identifies industries where the observed allocation of labour cannot be explained by aggregate scale effects alone, but instead reflects a genuine concentration of productive capabilities or specialisation relative to the structure of the global economy.

\begin{figure}[htbp]
    \centering    
    \caption{Comparison between ICA and RCA diversification ranking in 2014.}
    \includegraphics[width=0.95\linewidth]{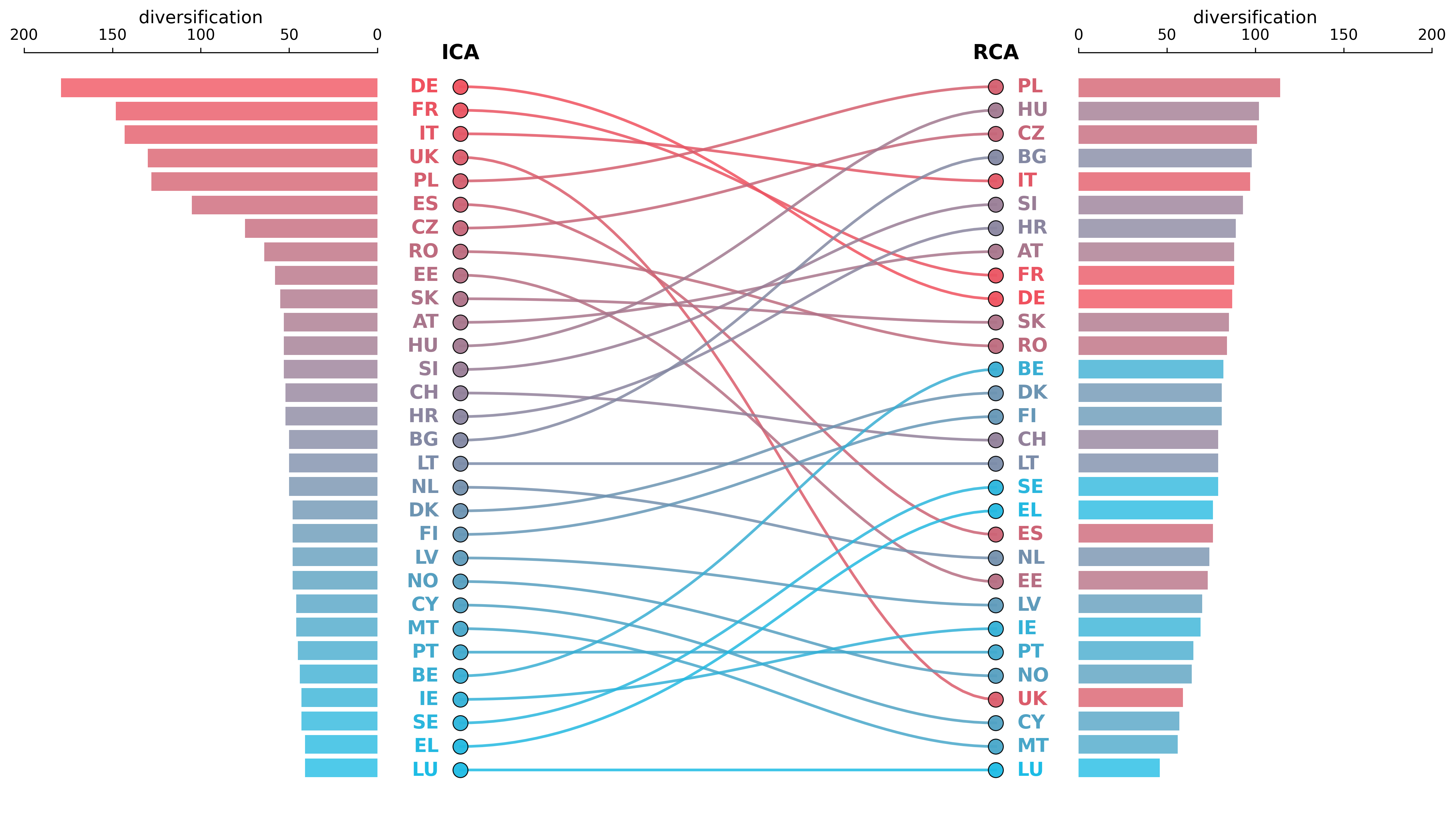}
    \begin{threeparttable}
        \begin{tablenotes}[flushleft]
            \footnotesize
            \item \textit{Note:} The two horizontal bar charts display each country's level of diversification as measured by the two comparative advantage indices.
            
        \end{tablenotes}
    \end{threeparttable}
    \label{fig:ranking_ica_rca}
\end{figure}

Figure~\ref{fig:ranking_ica_rca} compares the diversification rankings obtained using the ICA and RCA frameworks. The results reveal substantial differences between the two measures: for instance, countries such as the UK and Spain display almost opposite levels of relevance depending on the metric adopted. In contrast, Poland, Hungary, the Czech Republic, and Bulgaria rank at the top under the RCA framework and also remain among the highest‑ranked countries when evaluated with ICA. These discrepancies stem from the uneven distribution of the labour force across sectors and countries. Such imbalances influence how each CA framework captures diversification and the underlying capabilities. Indeed, when examining the distribution of diversification directly, it becomes evident that the RCA tends to produce a more uniform and less differentiated picture, whereas the ICA preserves stronger heterogeneity across countries.

\bigskip 

\subsection{Making Fitness comparable over time}\label{sec:fitscale}
The EFC algorithm computes a fixed point to determine the actual values of the Fitness and Complexity indices for each year independently. While the rankings of the indexes have a clear interpretation, their numerical values are relative to a scale that in unknown in the original formulation of the algorithm. What is the meaning of a Fitness $F_c = 5$? That depends on the scale we are using to measure the index. Such scale, if not fixed by the user, depends on the density of the input matrix and on the distribution of the points among rows and columns. Such quantities change over time and, with them, the reference scale of the values changes as well. Consequently, in principle, it is not possible to compare the index values across different years due to the dynamic changes in the bipartite networks. Negative changes of a country's Fitness may not reflect an actual loss of capabilities or competitiveness but a change in $M_{c,i}$ such as other countries gaining new products and changing the total density of the matrix. 

To address this issue, we employ a technique, the "dummy trick", to fix a constant reference scale for the numerical values as elaborated in~\citep{mazzilli2024equivalence}. We add to $M_{c,i}$ an additional row of all ones (the "dummy" row) which represents a fictitious country specialised in all industries. We rescale the Fitness values dividing all of them by the Fitness of the 'ideal' country, that is now fixed to $1$. Thus, in this version of the algorithm all $F_c < 1$ for all real countries. The scale of reference is fixed by the 'ideal' country that would export every product, and Fitness values indicate how close each country is to it. This approach establishes a fixed reference scale over time, enabling the comparison of Fitness and Complexity values across different years.

\end{document}